%% file: main.tex
\documentclass[sigconf]{acmart}
\AtBeginDocument{%
  \providecommand\BibTeX{{%
    \normalfont B\kern-0.5em{\scshape i\kern-0.25em b}\kern-0.8em\TeX}}}

\copyrightyear{2024}
\acmYear{2024}
\setcopyright{rightsretained}
\acmConference[CHI '24]{Proceedings of the CHI Conference on Human Factors in Computing Systems}{May 11--16, 2024}{Honolulu, HI, USA}
\acmBooktitle{Proceedings of the CHI Conference on Human Factors in Computing Systems (CHI '24), May 11--16, 2024, Honolulu, HI, USA}
\acmDOI{10.1145/3613904.3642939}
\acmISBN{979-8-4007-0330-0/24/05}

\acmConference[CHI 2024]{The ACM CHI Conference on Human Factors in Computing Systems}{11 May - 16 May, 2024}{Hawai'i, USA}
\acmBooktitle{CHI 2024: The ACM CHI Conference on Human Factors in Computing Systems, May 11 - 16, 2024, Hawai'i, USA}

\author{Zhe Liu$^{1,2}$,Chunyang Chen$^3$, Junjie Wang$^{1,2,*}$, Mengzhuo Chen$^{1,2}$, Boyu Wu$^{2,4}$, Yuekai Huang$^{1,2}$, \\ Jun Hu$^{1,2}$, Qing Wang$^{1,2,5,*}$}
\affiliation{
  \position{$^1$State Key Laboratory of Intelligent Game, Beijing, China}
  \department{Institute of Software Chinese Academy of Sciences, Beijing, China; \\
  $^2$University of Chinese Academy of Sciences, Beijing, China; $^*$Corresponding author;\\
  $^3$Technical University of Munich, Munich, Germany; $^4$DiDi Global Inc;
  }\city{$^5$Science \& Technology on Integrated Information System Laboratory, Beijing}
  \country{China}
}
\email{liuzhe181@mails.ucas.ac.cn, Chunyang.chen@monash.edu, junjie@iscas.ac.cn, wq@iscas.ac.cn}

\input{macro}

\begin{document}

\title{Unblind Text Inputs: Predicting Hint-text of Text Input in Mobile Apps via LLM}

\begin{abstract}
Mobile apps have become indispensable for accessing and participating in various environments, especially for low-vision users. Users with visual impairments can use screen readers to read the content of each screen and understand the content that needs to be operated. Screen readers need to read the hint-text attribute in the text input component to remind visually impaired users what to fill in. Unfortunately, based on our analysis of 4,501 Android apps with text inputs, over 76\% of them are missing hint-text. These issues are mostly caused by developers' lack of awareness when considering visually impaired individuals. To overcome these challenges, we developed an LLM-based hint-text generation model called {\tool}, which analyzes the GUI information of input components and uses in-context learning to generate the hint-text. To ensure the quality of hint-text generation, we further designed a feedback-based inspection mechanism to further adjust hint-text.
The automated experiments demonstrate the high BLEU and a user study further confirms its usefulness.
{\tool} can not only help visually impaired individuals, but also help ordinary people understand the requirements of input components.
{\tool} demo video: \url{https://youtu.be/FWgfcctRbfI}.

\end{abstract}


\maketitle

\input{sec/introduction}

\input{sec/related}

\input{sec/background}
\input{sec/motivation}
\input{sec/approach}

\input{sec/effectiveness}

\input{sec/evaluation}

\input{sec/discussion}

\input{sec/conclusion}

\begin{acks}
This work was supported by the National Natural Science Foundation of China Grant No.62232016, No.62072442 and No.62272445, Youth Innovation Promotion Association Chinese Academy of Sciences, Basic Research Program of ISCAS Grant No. ISCAS-JCZD-202304, and Major Program of ISCAS Grant No. ISCAS-ZD-202302.
\end{acks}

\bibliographystyle{ACM-Reference-Format}

\bibliography{reference}

\end{document}

%% file: macro.tex
\usepackage{mathtools}
\usepackage{soul}
\usepackage{float}
\usepackage{multirow}
\usepackage{epstopdf}
\usepackage{hyperref}
\usepackage{listings}
\usepackage{fancybox}
\usepackage{graphicx}
\usepackage{subcaption}

\usepackage{amsmath,bm}

\usepackage{array}
\usepackage{amsmath}
\usepackage{centernot}
\usepackage{xspace}
\usepackage{url}
\usepackage{verbatim}
\usepackage{wrapfig}
\usepackage{tabularx}
\usepackage{bbding}
\usepackage{pifont}
\usepackage{wasysym}
\usepackage{amssymb}

\newcommand{\tool}{\texttt{HintDroid}}

%% file: sec/introduction.tex
\section{Introduction}
\label{sec_introduction}
In the era of burgeoning mobile device development, mobile applications have evolved into indispensable tools for accessing and engaging within diverse environments, offering unparalleled convenience, particularly for individuals with low vision.
According to statistics from Google Play~\cite{GooglePlay} and App Store~\cite{AppStore}, over 4 million apps are widely used for various tasks such as entertainment, financial services, reading, shopping, banking, and chatting. However, many app developers often don't develop apps according to accessibility standards, which directly results in disabled people being unable to access the functionalities of many apps. This is especially true for visually impaired users, who need rich Graphical User Interface (GUI/UI) hints to understand the functionality. According to statistics from the World Health Organization (WHO)~\cite{Blindness-and-vision-impairment}, at least 2.2 billion people have near or distant vision impairment. In at least 1 billion of these, vision impairment could have been prevented or is yet to be addressed.~\cite{Blindness-and-vision-impairment}. This requires smart devices and mobile apps to provide convenience for them. Therefore, how to make mobile apps better serve their eyes is a social justice issue that the whole society needs to pay attention to~\cite{ladner2015design,Blindness-and-vision-impairment}.

\input{figure/Difference}

\input{figure/approach-example}

To facilitate interaction between visually impaired users and apps, an increasing number of mobile devices support screen reader apps~\cite{GoogleTalkBack,VoiceOver}. 
There are also many companies that have developed relevant screen readers for different disabled users to help them understand and use the functionalities of the app. The screen reader will recognize the ``\textit{text}'' or \textit{``content description''} of the \textit{text component}, ``\textit{label}'' of the \textit{image component} and the ``\textit{hint-text}'' of the \textit{text input component}, and read it aloud for visually impaired users. This is very effective for visually impaired users, who can understand the app's UI through screen readers and use their functionalities.
As shown in Figure \ref{fig:Difference}, \textit{hint-text} is different from \textit{label} and \textit{content description}. \textit{Label} is usually used to briefly describe image components as in Figure \ref{fig:Difference} (a), \textit{content description} provides the overview of related input components as in Figure \ref{fig:Difference} (b), while \textit{hint-text} goes a step further to explain the input requirements and help users understand what should input into the component (the red rectangles in Figure \ref{fig:Difference} indicate the example hint-text).
Google developer accessibility guideline~\cite{AndroidDeveloperAccessibility} requires developers to provide hint-text for input components, especially input components that lack content description. ``It's helpful to show text in the element itself, in addition to making this hint-text available to screen readers''~\cite{PrinciplesAccessibility}.

Despite the guideline, the real-world practice doesn't work like that. 
According to our observation on 4,950 popular apps from Google Play (in Section \ref{sec_motivation}), about 91\% of them have text input components, yet as high as 76\% of the input components have issues of missing \textit{hint-text} as shown in Figure \ref{fig:input-example} (a). 
Without \textit{hint-text}, the screen reader cannot obtain information about the text input component, which can cause it to skip the component. 
In addition, even if some input components have the hint-text, it can be simple and lacks practical meaning as in Figure \ref{fig:input-example} (b), which can't help people understand the text input requirements.
This also leads to disabled people being unable to use the functionalities provided by the app, which still affects their access to the functionality of this page.

\input{figure/input-example}

Therefore, it calls for automatic support for the generation of hint-text, yet to the best of our knowledge there is no existing study tackling this problem. 
The most relevant work is the generation of missing \textit{labels} for image components with deep learning techniques to improve the accessibility of apps~\cite{chen2020unblind,chen2022towards,zhang2021screen}. 
However, the approaches for generating \textit{labels} mainly involve the understanding of icons, while the generation of \textit{hint-text} not only requires more thorough viewpoints of the whole GUI page but also calls for deeper comprehension of the related information, as shown in Figure \ref{fig:Difference}.
Taken in this sense, this work will develop advanced techniques for generating the meaningful \textit{hint-text} as shown in \ref{fig:input-example} (c), which can help visually impaired users successfully fill in the correct input.

In order to overcome these challenges, inspired by the excellent performance of the Large Language Model (LLM) in natural language processing tasks~\cite{schulman2022chatgpt,chowdhery2022palm,zhang2022opt,brown2020GPT3,chen2020big,liu2023make}. We propose {\tool} for automated \textit{hint-text} generation based on LLM and GUI page information. Given the view hierarchy file of the GUI page, we first extract the text input component and the GUI entity information of the page, and then design GUI prompts to enable LLM to understand the text input context. In order to facilitate the LLM to better understand our task, we use in-context learning and construct an example database with hint-text and its corresponding input component information.
Then we design a retrieval-based example selection method to construct an in-context learning prompt. Combining the above two prompts, LLM will output hint-text and the input content generated based on it. To ensure the quality of hint-text, we use input content as a bridge to evaluate the generated hint-text, and extract feedback information by checking whether the input content can trigger the next GUI page, then construct a feedback prompt to further let LLM adjust the hint-text.

We evaluate the effectiveness of {\tool} on 2,659 text input components with hint-text involving 753 popular apps in Google Play (the largest repository of Android Apps). 
Results show that {\tool} can achieve 83\% BLEU@1, 77\% BLEU@2, 73\% BLEU@3, 66\% BLEU@4, 67\% METEOR, 63\% ROUGE and 62\% CIDEr with the generated \textit{hint-text}. 
Compared with 12 common-used and state-of-the-art baselines, {\tool} can achieve more than 82\% boost in the exact match compared with the best baseline.
In order to further understand the role of each module and sub-module of the approach, we conduct ablation experiments to further demonstrate its effectiveness.
We further carry out a user study to evaluate its usefulness in assisting visually impaired users, with 33 apps from Google Play. 
Results show that the participants with {\tool} fill in 152\% more correct input, cover 66\% more states and 77\% more activities within 139\% less time, compared with those without using {\tool}. 
The experimental results and feedback from these participants confirm the usefulness of our {\tool}.

The contributions of this paper are as follows:

\begin{itemize}

\item To our best knowledge, this is the first work to automatically predict the \textit{hint-text} of text input components for enhancing app accessibility\footnote{We release the source code, experiment detail, and demo videos of our {\tool} in https://github.com/franklin/HintDroid. The demo video link is \url{https://youtu.be/FWgfcctRbfI}. \label{github}}. We hope this work can invoke the community's attention to maintaining the accessibility of mobile apps from the viewpoint of hint-text.

\item An empirical study for investigating how well the text input components of current apps support the accessibility for users with vision impairment, which motivates this study as well as follow-up researches.

\item Large-scale evaluation on real-world mobile apps with promising results, and a user study demonstrating {\tool}'s usefulness in assisting visually impaired users successfully fill in the correct input.
\end{itemize}

%% file: figure/Difference.tex
\begin{figure}[htb]
\centering
% \vspace{0.05in}
\includegraphics[width=8.3cm]{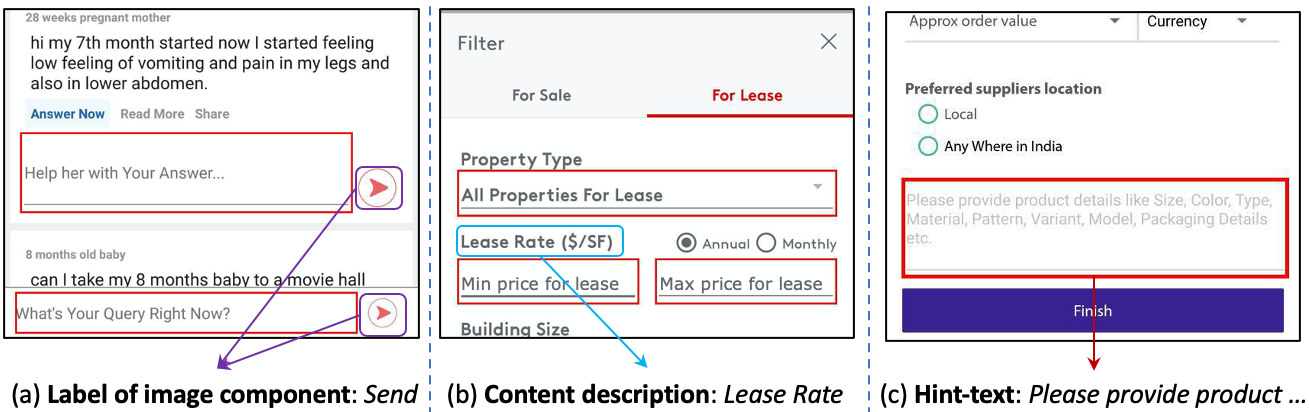}
\caption{Examples of differences between hint-text, label and context description. (a) Label is used to briefly describe image components. (b) Content description provides an overview of related input components. (c) Hint-text further explains the input requirements.}
\vspace{-0.1in}
\label{fig:Difference}
% \vspace{-0.05in}
\end{figure}

%% file: figure/approach-example.tex
\begin{figure*}[htb]
\centering
\vspace{0.05in}
\includegraphics[width=17.3cm]{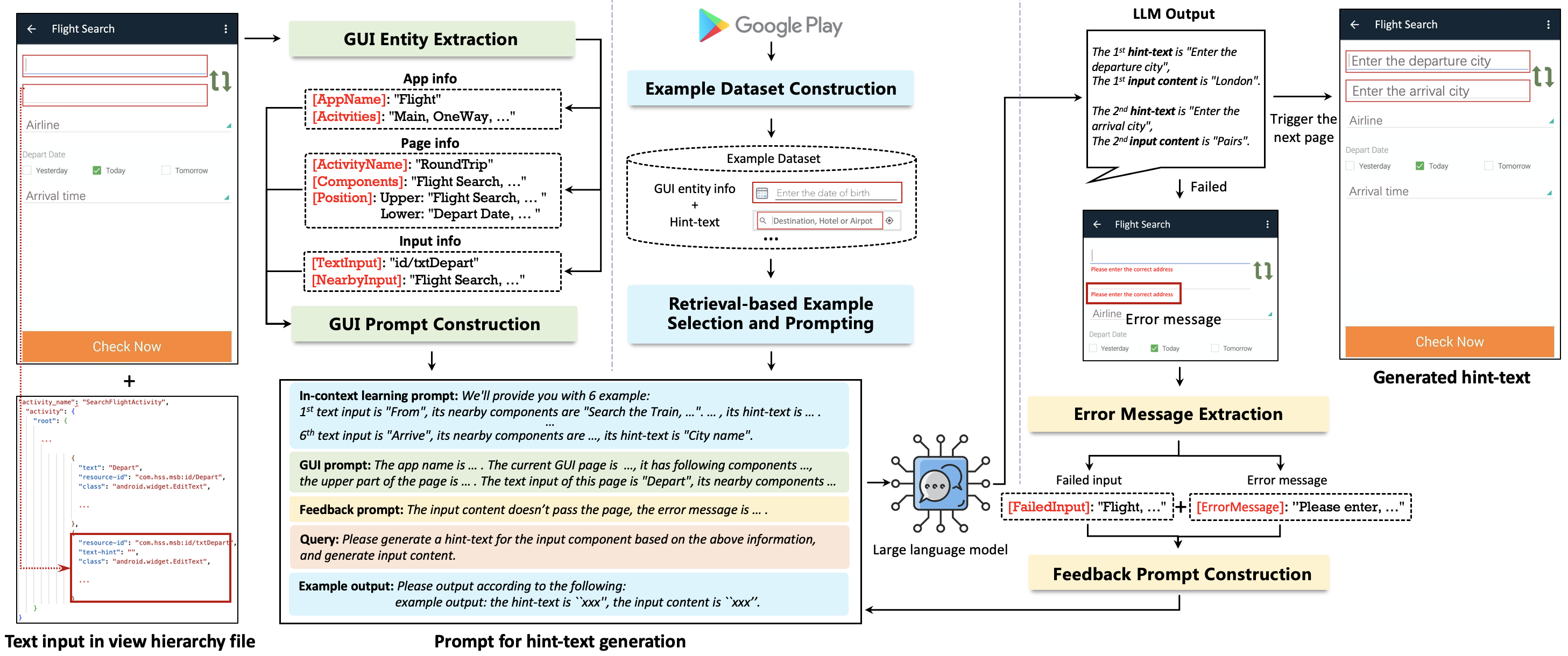}
\caption{Workflow of our {\tool}: It extracts GUI entity information from the view hierarchy file of the GUI page and constructs a GUI prompt that helps LLM understand the context. To facilitate LLM's better understanding of the task, {\tool} uses a retrieval-based example selection method to construct the in-context learning prompts. It also uses input content as a bridge to evaluate the generated hint-text and extracts feedback information by checking whether the input content can trigger the next GUI page.}
\vspace{-0.1in}
\label{fig:approach-example}
% \vspace{-0.05in}
\end{figure*}

%% file: figure/input-example.tex
\begin{figure}[htb]
\centering
% \vspace{0.1in}
\includegraphics[width=8.3cm]{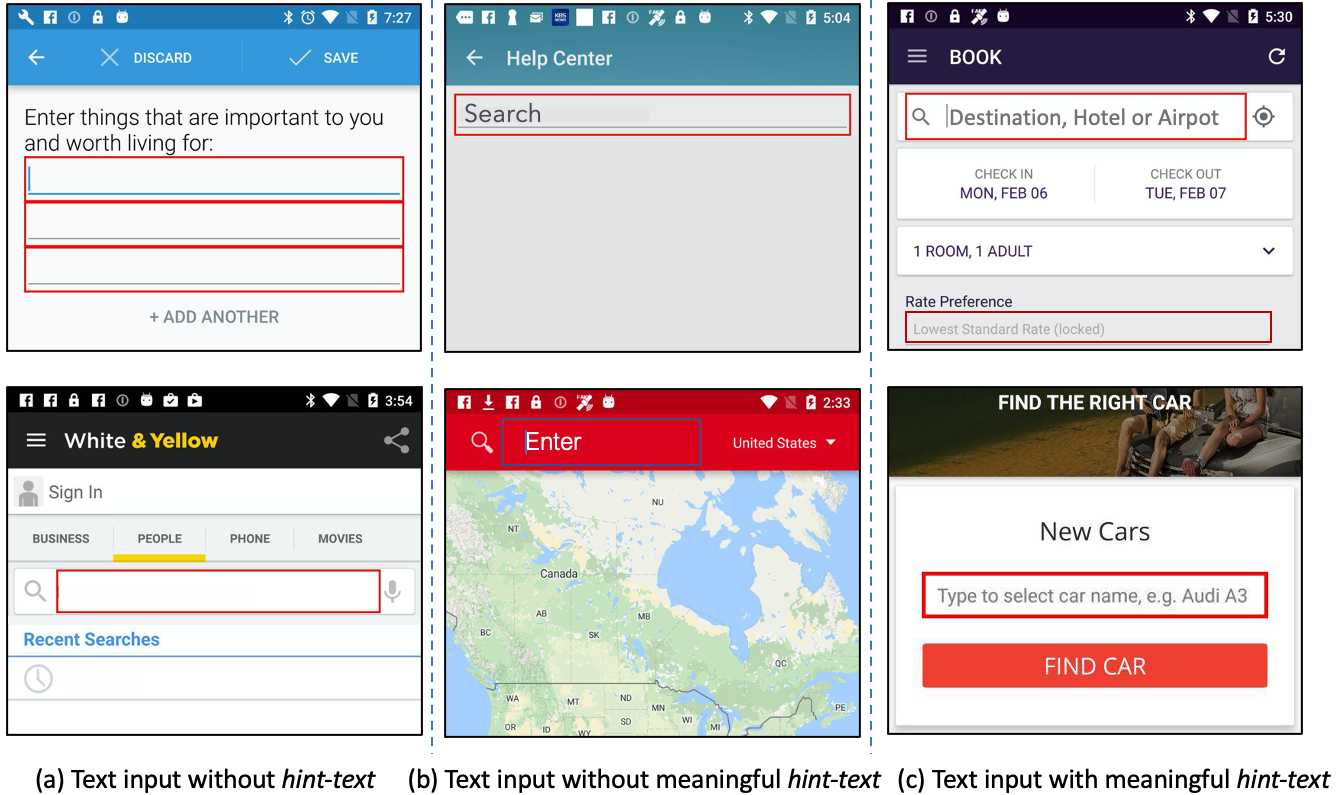}
 \caption{Example of the text input component without/with hint-text.(a) These inputs have issues of missing hint-text. (b) Hint-text lacks practical meaning. (c) Hint-text can help visually impaired users successfully fill in the correct input.}
\vspace{-0.1in}
\label{fig:input-example}
\vspace{-0.1in}
\end{figure}

%% file: sec/related.tex
\section{Related Work}
\label{sec_related}
With the development of mobile applications, more and more companies (Google, Apple) recognize the need to improve the accessibility of apps and have introduced guidelines for app developers and designers~\cite{AppleAccessibility,AndroidDeveloperAccessibility,PrinciplesAccessibility,GoogleAccessibility}, including basic principles for accessibility design. The accessibility of apps is crucial for users to use the app normally, especially for visually impaired users and other disabled users~\cite{power2012guidelines,vendome2019can,morris2014blind,jain2021smartphone,shaik2010design,zhao2019seeingvr,guinness2018caption,neat2019scene,salehnamadi2022groundhog}. Missing hint-text in the input text component is a common accessibility issue that can affect users' understanding of input requirements, especially for visually impaired users, as screen readers need to provide input requirements to visually impaired users by reading hint-text fields~\cite{VoiceOver,GoogleTalkBack,PrinciplesAccessibility,AndroidDeveloperAccessibility}. Therefore, this paper focuses on the accessibility issue of text input components. Our {\tool} can automatically complete the hint-text of the input components, helping visually impaired people understand the text input components and successfully fill in the correct input.

\subsection{App Accessibility for Visually Impaired Users}
Within the domain of Human-Computer Interaction, researchers have extensively delved into accessibility challenges prevalent in various categories of small-scale mobile applications~\cite{khan2021insight,park2014toward,yan2019current,liu2020owl,liu2022Nighthawk,liu2022Guided,inal2020perspectives}, spanning domains like health~\cite{milne2014accessibility,daihua2015accessibility,jones2018mobile,ramey2019apps,wang2012measurement}, smart cities~\cite{mora2017comprehensive,wu2020mobile,visvizi2019sustainable}, and government engagement~\cite{king2016government,akgul2022evaluating}. While these works scrutinize distinct facets of accessibility, a recurring theme has been the conspicuous absence of descriptions for image-based components~\cite{ross2018examining,kreiss2022context,tian2010computer,morris2016most}. The gravity of this issue has been explicitly recognized across studies. Notably, Park et al.~\cite{park2014toward} conducted an evaluation that assigned both severity and frequency ratings to different accessibility errors, with missing labels emerging as the most severe among various issues. Kane et al.~\cite{kane2009freedom} conducted an investigation into mobile device adoption and accessibility for users with visual and motor disabilities. Ross et al.~\cite{ross2018examining} conducted a comprehensive analysis of image-based button labeling within a relatively large corpus of Android apps, pinpointing prevalent labeling deficiencies. 
Based on the above analysis of label missing issues, Chen et al.~\cite{chen2020unblind} proposed LabelDroid, and employed deep-learning techniques to train a model on a dataset of existing icons with labels to automatically generate labels for visually similar, unlabeled icons.
To further improve the performance of label generation, Mehralian et al.~\cite{mehralian2021data} considered more GUI information 
and proposed a context-aware label generation approach, COALA, that incorporated several sources of information from the icon in generating accurate labels. 
The above studies have conducted in-depth research on the accessibility issues of image components. In addition, statistical data shows that the accessibility issues associated with text input are also serious, but this issue has received relatively little attention in existing research. So our study not only presents the most expansive scrutiny of text input components but also introduces an LLM-based solution to generate hint-text.

Several works also focused on detecting and rectifying accessibility gaps, particularly for users with visual impairments~\cite{sierra2012designing,alshayban2020accessibility,leporini2012interacting,chiou2023bagel,ballantyne2018study,ahmetovic2015zebra,salehnamadi2021latte,acosta2021accessibility,liu2023ex,liu2022navidroid}. Eler et al.~\cite{eler2018automated} proposed an automated test generation model to dynamically evaluate mobile apps. Salehnamadi et al. ~\cite{salehnamadi2021latte} designed a high-fidelity form of accessibility testing for Android apps, Latte, that automatically reused tests written to evaluate an app’s functional correctness to assess its accessibility as well. Zhang et al.~\cite{zhang2018robust} leveraged crowd-sourcing to annotate GUI elements devoid of original content descriptions. 
Although these researches help enhance mobile accessibility, they still can't fix the issue of missing hint-text. This also further motivates us to design an automated method to generate hint-text that satisfies contextual semantics.

\subsection{GUI Understanding and Intelligent Interactions for Visually Impaired Users}
In order to help visually impaired individuals understand the meaning of GUI pages and components, researchers have attempted to use computational vision technology for GUI modeling and semantic understanding of GUI pages~\cite{deka2017rico,kumar2013webzeitgeist,zhang2021screen,burns2022interactive,yeh2009sikuli,feiz2022understanding,johns2023interactive,gur2022understanding,bai2021uibert,wu2023never,fu2021understanding,ang2022learning}. Schoop et al.~\cite{schoop2022predicting} designed a novel system that models the perceived tappability of mobile UI elements with a vision-based deep neural network and helps provide design insights with dataset-level and instance-level explanations of model predictions. He et al.~\cite{he2021actionbert} designed a new pre-trained UI representation model, ActionBert, aimed at utilizing visual, linguistic, and domain-specific features from user interaction traces to pre-train the universal feature representation of UI and its components. 
Although these studies can help users understand the GUI information of pages, they have not identified and understood the relevant information of text input components.
Considering that visually impaired users cannot obtain GUI information on the page through their eyes, it is difficult to determine input requirements. To address the above challenges, this paper repairs the hint-text missing issue in input components and helps visually impaired users understand the input components.

The portability of mobile devices has led to an increasing number of visually impaired people using smartphones for daily life~\cite{kientz2006s,salehnamadi2022groundhog,jain2021smartphone,rodrigues2015getting}. Researchers have explored innovations in alternative sensory modalities like speech systems~\cite{paek2007improving,azenkot2013exploring}, auditory~\cite{brewster2002overcoming,gori2014impairment}, and multimodal interaction~\cite{brewster2007tactile}, etc. It has opened a new era for accessible usage of the smartphone for visually impaired people~\cite{nah2017editorial,damaceno2018mobile}. To assist visually impaired people in filling in the input content, researchers proposed some input solutions rely on indicating Braille mappings to enter characters~\cite{azenkot2012input,luna2023text,gaines2023flextype,mascetti2012typeinbraille,komninos2023review,oliveira2011brailletype,gaines2018exploring,bonner2010no,tinwala2010eyes,alnfiai2016singletapbraille}.
While input methods to assist mobile text entry have been extensively studied in the recent literature~\cite{liu2023fill,liu2017automatic}, text entry research has focused much less on the needs of persons with vision problems. For visually impaired people, how to use the text input functionality of an app is a challenging task. They not only need to understand the intent of the input component, but also fill in the correct input.

\subsection{LLM Usage in Human-computer Interaction}
Recently, the great success of pre-trained Large Language Models~\cite{schulman2022chatgpt,chowdhery2022palm,zhang2022opt,brown2020GPT3,chen2020big} in a variety of NLP tasks. Considering the powerful performance of LLM, researchers have successfully leveraged it to solve various tasks in the field of 
human-machine interaction and software engineering~\cite{56retrievalPromptSelection,14conformanceTesting,1itigerIssueTitle,40circleContinualRepair,35vulnerabilityRepair}.
Supported by code naturalness~\cite{hindle2016naturalness}, researchers applied the LLMs to code writing in different programming languages~\cite{feng2020codebert,xu2022systematic}. 
A similar work QTypist~\cite{liu2023fill} leveraged the LLM to generate the text inputs for triggering the next GUI page in order to improve the testing coverage of mobile testing.
Different from its sole focus on text input generation to boost existing GUI testing tools, {\tool} is designed for generating the hint-text of text input component, which can help visually impaired users fill in the correct input.

LLMs were also successfully applied in research related to the HCI community ~\cite{chung2022talebrush,jiang2022promptmaker,jiang2022discovering,kim2022stylette,lee2022promptiverse,liu2022will,wang2023enabling}.
Stylette~\cite{kim2022stylette} allowed users to modify web designs with language commands and used LLMs to infer the corresponding CSS properties. Lee et al.~\cite{lee2022coauthor} presented CoAuthor, a dataset designed to reveal GPT-3's capabilities in assisting creative and argumentative writing. Othman et al.~\cite{othman2023fostering} proposed an automated accessibility issue repair method based on LLM, which utilizes LLM to repair website accessibility issues for the first time. Wang et al.~\cite{wang2023enabling} investigated the feasibility of enabling versatile conversational interactions with mobile UIs using a single LLM and designed prompting techniques to adapt an LLM to mobile UIs.  
These researches on LLM also inspires us to use LLM knowledge to understand UI and generate hint text.
Unlike them, this paper addresses the challenge of missing hint-text that cannot be fixed by current accessibility repair tools. To our knowledge, it is the first time to propose using LLM's automation to repair hint-text missing issues, helping users further understand input requirements.

%% file: sec/background.tex
\section{Android accessibility background}
\label{sec_background}
In this section, we first introduce Android-related terminology and Android screen reader. This background helps to support the concise and clear terminology used throughout our analysis.

\textbf{Android Text Input Components.} When developers develop UI, the Android platform provides them with many different types of UI components~\cite{AndroidDeveloperAccessibility}, such as TextView, ImageView, EditText, Button, and so on. EditText is a user interface element for entering and modifying text. When you define an EditText component, you must specify the \textit{R.styleable.TextView\_InputType} attribute. The attribute   ``hint'' is mainly used to display prompts for input requirements. For text input components, Google developer accessibility guideline~\cite{AndroidDeveloperAccessibility} requires developers to provide the \textit{Android: hint} attribute to provide input requirements for users.

\textbf{Android Screen Readers.} Based on the WebAIM~\cite{ScreenSurvey}, a significant 95.1\% of visually impaired respondents relied on smartphone screen readers. Google and Apple, as major industry players, shape mobile technology and apps. For vision-impaired users, Google's TalkBack~\cite{GoogleTalkBack} and Apple's VoiceOver~\cite{VoiceOver} provide crucial accessibility, enabling mobile app engagement.
Focusing on Android apps, TalkBack is a cornerstone accessibility service. Pre-installed on many Android devices, it empowers blind and visually impaired users. TalkBack narrates blocks of text, alerts users to interactive elements, like buttons, and extends beyond text, engaging with apps through intricate gestures. It also offers local and global context menus, allowing tailored interaction experiences and global setting adjustments.
A silent ally for those navigating digital realms with different senses, it bridges the gap between user intent and device response, enhancing mobile app experiences universally. TalkBack typically obtains label and hint-text from apps to provide services to users, so providing label and hint-text in mobile applications is crucial for helping users understand the app.

%% file: sec/motivation.tex
\section{motivational study}
\label{sec_motivation}

\input{figure/motivation-1}

This paper focuses on proposing an automated approach to automatically generate the hint-text for text input components. 
Compared to the purpose of input content generation, hint-text focuses more on helping users or visually impaired individuals understand the needs of input components from a human cognitive perspective.
To understand the distribution of text input components and hint-text missing issues in popular apps, we conduct a motivational study to explore the potential impact of hint-text missing issues on people with disabilities.

\subsection{Data Collection}
\label{sec_motivation_Data Collection}
In order to assess the support of popular mobile apps for vision impairment users, we randomly crawl 4,950 apps from 33 categories (150 apps per category) from Google Play~\cite{GooglePlay}, all of which were updated between December 2022 and June 2023, with installations ranging from 1K to 100M. We use the Application Explorer~\cite{chen2018ui} (ensuring that each app can run normally on the browser) to automatically explore different screens in the app through various operations such as clicking, editing, and scrolling. During the exploration process, we get screenshots of the app GUI and their view hierarchy file (runtime front-end code file), which identifies the type of each element (such as EditText, TextView), coordinates in the screenshot, content description, and other metadata. After removing all duplicate screenshots, we ultimately collected 73,286 GUI screenshots and their corresponding View hierarchy files from 4,950 apps. In the collected data, 45,803 (63\%) screenshots from 4,501 apps (91\%) included text input components, which formed the dataset we analyzed in this study.

\subsection{Current Status of Text Input Component Hint in Mobile Applications}
\label{sec_motivation_Result}
Statistical results show that out of 4,501 apps with text input components, 3,398 (76\%) of them are without hint-text. Among all 45,803 screens, 30,226 (66\%) had at least one text input without explicit hint-text content. This means that more than half of the text input components do not provide hint-text to users. These statistical data confirm the severity of the issue of missing hint-text, which may seriously hinder visually impaired users from using the mobile app. Then, we further analyze the issue of missing hint-text in the text input component of different categories of mobile apps. The statistical results show that the issue of hint-text missing widely exists in different types of mobile apps, and some categories of hint-text missing problems are severe. As shown in Figure \ref{fig:motivation-1} (a), more than 80\% of the hint-text of 18 categories of apps are missing. The missing rate of hint-text in \textit{Tools}, \textit{Shopping}, \textit{Book and Reference} and \textit{Travel and local} categories, which commonly used in the daily lives of visually impaired individuals exceeds 90\%, greatly affecting their daily use.

According to Google Play's download statistics standards~\cite{GooglePlay}, We further analyze the hint-text missing rate for apps with different download numbers (popularity). We find that regardless of the number of downloads, the apps have a serious hint-text missing rate, of 71\%-89\%. Meanwhile, we also find that popular apps with high downloads also have serious issues of missing hint-text, i.e., apps with 100K-100M downloads have similar missing rates, with an average missing rate of 72\%.
Such hint-text missing issues with popular apps may have a greater negative impact, as these apps have a larger audience. These findings confirm the severity of the issue of missing hint-text, therefore an automated method is needed to generate missing hint-text for text input components.

%% file: figure/motivation-1.tex
\begin{figure*}[htb]
\centering
\vspace{0.05in}
\includegraphics[width=17.3cm]{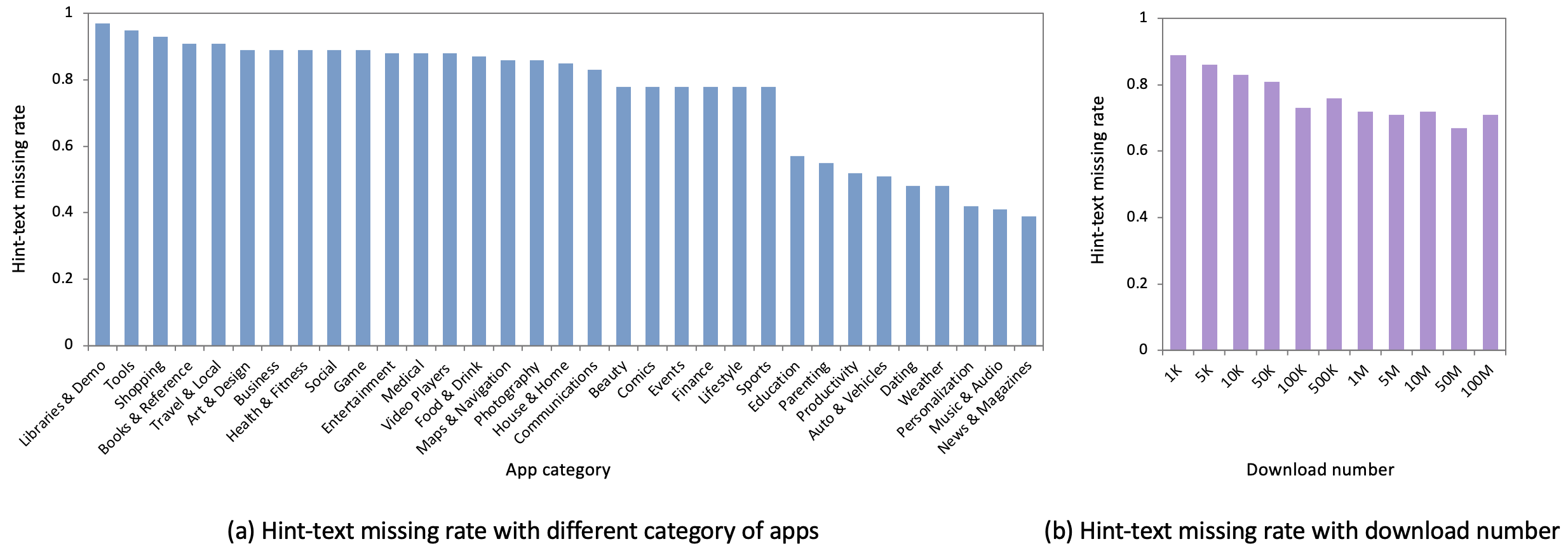}
\caption{Statistical results of hint-text missing rate. More than 80\% of the hint-text of 18 categories of apps are missing.}
\vspace{-0.1in}
\label{fig:motivation-1}
\vspace{-0.05in}
\end{figure*}

%% file: sec/approach.tex
\input{figure/overview}

\section{Approach}
\label{sec_approach}
This paper proposes {\tool}, which uses LLM to automatically generate the hint-text of text input components to facilitate the screen reader of disabled users in better understanding the inherent meaning of the input components.
Figure \ref{fig:overview} presents the overview of {\tool}, which consists of three main modules. 
\textbf{\textit{First}}, given the view hierarchy file of the text input page of the Android application, the GUI entity extraction module extracts GUI information related to the text input component and the contextual information of its nearby components. Then, based on these extracted GUI information, {\tool} generates LLM understandable GUI prompts for hint-text generation (Section \ref{subsec_approach_GUI_extraction}).
\textbf{\textit{Second}}, based on the GUI prompt, the information retrieval-based prompt enhancement module automatically selects examples that are most similar to the current input scenario, and constructs an in-context learning prompt. This prompt uses in-context learning to help LLM better understand hint-text generation tasks, thereby enhancing LLM's hint-text generation performance (Section \ref{subsec_approach_in-context}).
\textbf{\textit{Third}}, with the above prompt, we instruct the LLM to generate not only the hint-text, but also the suggested input content that aligns with the hint-text.
We then input the content into the component and check whether it triggers the next page. 
If it doesn't trigger the next page, we assume the generated hint-text may be inappropriate, and will provide the failed case as the feedback prompt to facilitate the LLM in adjusting the answer. 
Additionally, when receiving inappropriate input content, certain text input components can also report the error message for indicating the expected input, and we also utilize this error message in the feedback for re-querying the LLM (Section \ref{subsec_approach_FeedBack}).

\subsection{GUI Entity Extraction and Prompting}
\label{subsec_approach_GUI_extraction}
The first module of our approach is to understand, analyze, organize, and extract the entities from the GUI page with the text input component. 
Although LLM is equipped with knowledge learned from large-scale training corpora and excels in various tasks, its performance may be significantly affected by the quality of its input, that is, whether the input can accurately describe what to ask.
Therefore, we design an approach to extract and organize the GUI information.

\input{tab/example-entity}

\subsubsection{\textbf{GUI Entity Extraction}}
\label{subsubsec_approach_GUI_Entity_Extraction}
Inspired by the screen reader's ability, we convert GUI information into natural language descriptions~\cite{GoogleTalkBack,VoiceOver,li2021screen2vec,zhang2021screen}.
We first extract the GUI entity information of the app, the GUI page currently used, and the components on the page, which helps LLM understand the GUI of the current page and the input requirements of the text input components. The app entity information is extracted from the \textit{AndroidMaincast.xml} file, while the other two types of entity information are extracted from the view hierarchy file, which can be obtained by UIAutomator~\cite{uiautomator}.
Table \ref{tab:example-entity} presents the summarized view of them.

\textbf{App Entity Information} provides the macro-level semantics of the app under testing, which facilitates the LLM to gain a general perspective about the functionalities of the app. The extracted information includes the app name and the name of all its activities. The app name can help LLM understand the type of app it belongs to, and the activity name can help LLM infer the current functional GUI page and the functional GUI page after entering the input content, thereby providing a basis for generating hint-text that conforms to contextual semantics. 

\textbf{Page GUI Entity Information} provides the semantics of the current GUI page during the interactive process, which facilitates the LLM to capture the current snapshot.  
We extract the activity name of the current GUI page, all the components represented by the ``text'' field or ``resource-id'' field (the first non-empty one in order), and the component position of the page.
For the position, inspired by the screen reader~\cite{GoogleTalkBack,VoiceOver}, we first obtain the coordinates of each component in order from top to bottom and from left to right, and the components whose ordinate is below the middle of the page are marked as lower, and the rest is marked as upper.

\textbf{Input Component Entity information} denotes the micro-level semantics of the GUI page, i.e., the inherent meaning of all its text input components, which facilitates the LLM in understanding the input requirements and the semantic information of its contextual components.
The extracted information includes ``text'' and ``resource-id'' fields (the first non-empty one in order).
To avoid the empty textual fields of a component, we also extract the information from nearby components to provide a more thorough perspective, which includes the ``text'' of parent node components and sibling node components.

\input{tab/example-prompt}

\subsubsection{\textbf{GUI Prompt Construction}}
\label{subsubsec_approach_GUI_Prompt_Generation}

With the extracted GUI information, we combine them into the GUI prompt that LLM can understand, as shown in Table \ref{tab:example-prompt}, i.e., <App information> + <GUI page information> + <input component formation> + <Query> + <Example output>.
Generally speaking, it first provides the app information, GUI page information and the input component information, then queries the LLM for the hint-text of the text input and its corresponding input content and gives it the example output as shown in Figure \ref{fig:prompt-example}. Due to the robustness of LLM, the generated prompt sentence does not need to fully follow the grammar. After inputting the GUI prompt, the LLM will return its recommended hint-text and its inferred input content.

\subsection{Enriching Prompt with Examples}
\label{subsec_approach_in-context}
It is usually difficult for LLM to perform well on domain-specific tasks as our hint-text generation, and a common practice would be employing the in-context learning~\cite{dong2022survey,min2022rethinking,shin2022effect} schema to boost the performance. 
It provides the LLM with examples to demonstrate what the instruction is to enable the LLM to better understand the task. 
Following the in-context learning schema, along with the GUI prompt for the text input as described in Section \ref{subsubsec_approach_GUI_Prompt_Generation}, we additionally provide the LLM with examples of the hint-text. 
To achieve this, we first build a basic example dataset of hint-text from the popular mobile apps in our motivational study. 
Research shows that the quality and relevance of examples can significantly affect the performance of LLM~\cite{dong2022survey,min2022rethinking,shin2022effect}. Therefore, based on the dataset we built, a retrieval-based example selection method (in Section \ref{subsubsec_approach_Retrieval}) is designed to select the most appropriate example according to the text input and its hint-text, which further enables the LLM to learn with pertinence. 

\subsubsection{\textbf{Example Dataset Construction}}
\label{subsubsec_approach_dataset_construction}
We collect the hint-text of text input components from Android apps in our motivational study and continuously build an example dataset that serves as the basis for in-context learning. 
For each data instance, as demonstrated in Table \ref{tab:example-prompt}, it records the GUI information of text input components and its hint-text, which enables us to select the most suitable examples and facilitate the LLM understanding of what the hint-text and the text input context like.

\textbf{Mining Hint-text from Android App.} 
First, we automatically crawl the view hierarchy file from the Android mobile apps in our motivational study. 
Then we use keyword matching to filter these related to the text input components (e.g., EditText) which have hint-text. 
In this way, we obtain 15,577 (45,803 - 30,226) view hierarchy file in Section \ref{sec_motivation} with text input components (all of them has the hint-text) and store them in the example dataset (there is no overlap with the evaluation datasets).
We then extract the GUI entity information of the text input component with the method in Section \ref{subsubsec_approach_GUI_Entity_Extraction}, and store it together with the hint-text.

\textbf{Enlarging the Dataset with Hint-text During Using.}
We also enrich the hint-text example dataset with the new hint-text which generates input content that truly triggers the next page (transition) during {\tool} runs on various apps. 
Specifically, for each generated hint-text, after running it in the mobile apps, we put the ones that generate input content that truly triggers the next page (transition) and its GUI information into the example dataset.

\subsubsection{\textbf{Retrieval-based Example Selection and In-context Learning}}
\label{subsubsec_approach_Retrieval}
The hint-text examples can provide intuitive guidance to the LLM in accomplishing a task, yet excessive examples might mislead the LLM and cause the performance to decline. 
Therefore, we design a retrieval-based example selection method to choose the most suitable examples (i.e., most similar to the input components) for LLM.

In detail, the similarity comparison is based on the GUI entity information of the text input components. 
We use Word2Vec (Lightweight word embedding method)~\cite{Word2Vec} to encode the context information of each input component into a 300-dimensional sentence embedding, and calculate the cosine similarity between the input component and each data instance in the example dataset. 
We choose the top-K data instance with the highest similarity score, and set K as 6 empirically. 
The selected data instances (i.e., examples) will be provided to the LLM in the format of GUI entity information and hint-text, as demonstrated in Table \ref{tab:example-prompt}.

\input{figure/prompt-example}

\subsection{Feedback Extraction and Prompting}
\label{subsec_approach_FeedBack}
Since the LLM cannot always generate the correct hint-text as we want, we ask it to generate not only the hint-text, but also the input content that aligns with the hint-text, and use the input content to help determine whether the hint-text is appropriate or not. 
The design idea of HintDroid is to role-play LLM as a low vision user, providing it with GUI context and hint-text to see if it can generate input content correctly. So, we optimize the hint-text by determining whether the generated input content can trigger the next page, rather than simply triggering the next page which is what QTypist~\cite{liu2023fill} did for software testing.
In detail, we enter the generated content into the text input component and check whether it triggers the next page (transition). 
If it doesn't trigger the next page (transition), we use the failed information to serve as the feedback and add it to the prompt to re-query the LLM.

\subsubsection{\textbf{Automated Input Content Checking}}
\label{subsubsec_approach_judgment}
We design an automated script to automatically input the content into components and perform operations. Specifically, LLM outputs the hint-text and input content of each text input component in a fixed format. We will recode it into an operation script in Android ADB format, such as \textit{adb shell input text ``xxx''}. Then we iterate through the components of the current page, identify the components that require further action after input is completed, and use the ADB command to perform the operation. After the operation is completed, we will check whether the situation detector has triggered the next page. Determine whether the page name and components have changed. 
For cases that failed trigger, we will proceed with the following feedback extraction.

\subsubsection{\textbf{Error Message Extraction}}
\label{subsubsec_approach_feed}
When one inputs an incorrect text into the component, they might report the error message, e.g., the app may alter the users that the password should contain letters and digits. 
The error message can further help LLM understand what the valid input should look like. 

We extract the error messages via differential analysis which compares the differences between the GUI page before and after inputting the text, and extracts the text field of the newly emerged components (e.g., a popup window) in the later GUI page, with examples shown in Figure \ref{fig:approach-example}.
We also record the text input which makes the error message happen, which can help the LLM to understand the reason behind it.

\subsubsection{\textbf{Feedback Prompt Construction}}
\label{subsubsec_approach_feed_prompt}
For cases where there is no successful page transfer, we combine the aforementioned information in Table \ref{tab:example-prompt} as the feedback prompt, i.e. <Feedback> + GUI prompt + <Feedback query> + <Example output>.
We design two types of <Feedback>. One is to tell the LLM that the current hint-text and input content are incorrect. The other is to provide the information dynamically generated by the text input component, which may make people more clear at once.
Note that not all input components may provide an error message, and for components without an error message, we display null in the prompt. Finally, we request LLM to re-optimize the hint-text based on the feedback prompt mentioned above.

\subsection{Implementation}
\label{subsec_approach_Implement}
We implement {\tool} based on the Turbo-3.5
which is released on the OpenAI website\footnote{\url{https://beta.openai.com/docs/models/chatgpt}}. 
It obtains the view hierarchy file of the current GUI page through UIAutomator~\cite{uiautomator} to extract GUI information of the text input components.
{\tool} can be integrated by the automated GUI exploration tool, which automatically extracts the GUI information and generates the hint-text.
When we obtain the generated hint-text from {\tool}, we further design an automated script to add its missing hint-text to the app. Specifically, as shown in Figure \ref{fig:implement}, given any app, the automated script consists of the following five steps. (1) We use Application Explorer~\cite {chen2018ui} to automatically run the app and get the view hierarchy files for each page. (2) We detect GUI pages (view hierarchy files) with missing hint-text based on text input components. (3) {\tool} automatically generates the corresponding hint-text. (4) Automatically decompile the APK file of the app and retrieve the code of text input. We design a script to automatically execute to decode the APK. (5) Automatically repackage APK files. We design a script to automatically execute to encode and package the modified code into APK, and complete the repair. Please note that if it is an open-source app, we can directly modify its source code and repackage it. So {\tool} is not a dynamic interactive process, it is a one-time offline job.
We also calculate that the average time for generating hint-text on each GUI page is 1.86 seconds.

\input{figure/implement}

%% file: figure/overview.tex
\begin{figure*}[htb]
\centering
\vspace{0.1in}
\includegraphics[width=17.3cm]{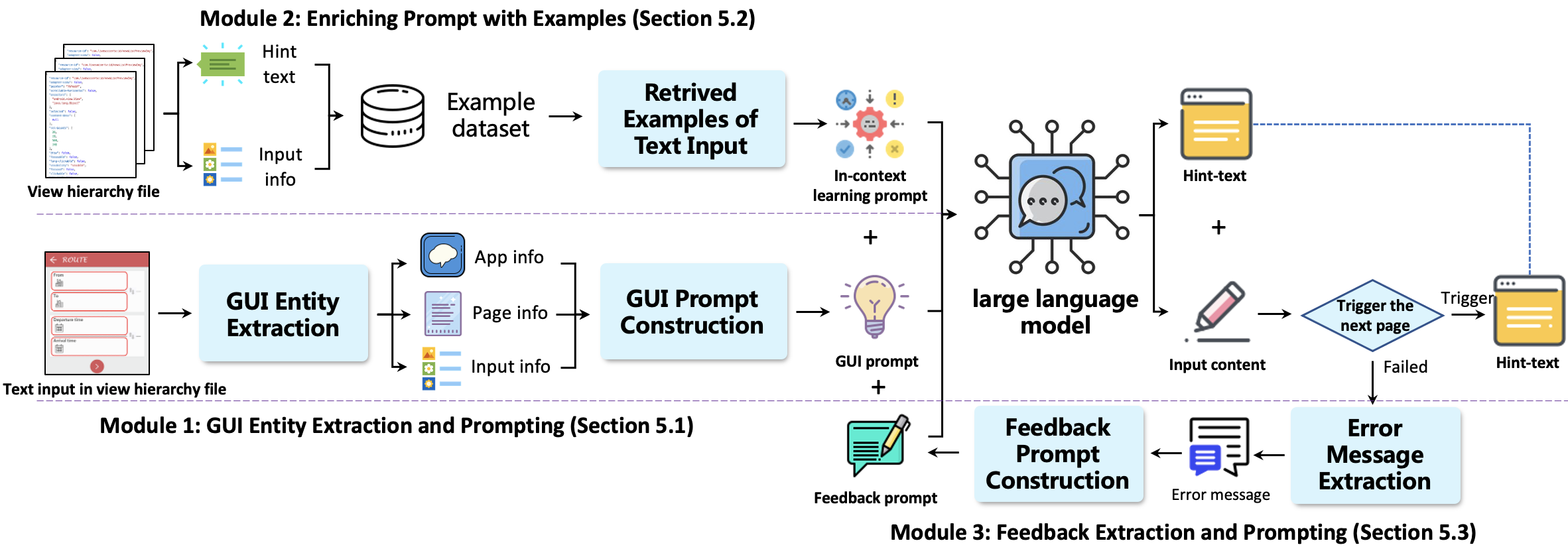}
\vspace{-0.1in}
\caption{Overview of {\tool}. {\tool} consists of three main modules: (1) Module 1 is used to extract the contextual GUI information of the text input and generate the GUI prompt. (2) Module 2 is used to construct the in-context learning prompt to improve the performance of LLM. (3) Module 3 further optimizes the generation results of hint-text through a feedback mechanism.}
\label{fig:overview}
\vspace{-0.1in}
\end{figure*}

%% file: tab/example-entity.tex
\begin{table*}[htb]
\vspace{0.1in}
\caption{The example of the GUI entity extraction. {\tool} extracts the GUI entity information of the app, the GUI page currently used, and the components on the page.}
\vspace{-0.1in}
\label{tab:example-entity}
\centering
\footnotesize
\begin{center}
\begin{tabular}{m{2.2cm}<{\centering} | m{1.5cm}<{} | m{5.6cm}<{} | m{6.1cm}<{}}
\toprule
\textbf{Type} & \textbf{Entity} & \textbf{Description}  & \textbf{Instantiation} \\
\midrule
\multirow{2}{2.2cm}{App information} & [AppName] & Name of the app under testing & [AppName]:``Flight'' \\
 & [Activities] & List of names for all activities of the app, obtained from \textit{AndroidManifest.xml} file & [Activities]: ``Main, OneWay, RoundTrip, ...'' \\
\midrule
\multirow{3}{2.2cm}{Page GUI information} & [ActivityName] & Activity name of the current GUI page & [ActivityName]: ``RoundTrip'' \\
 & [Component] & List of all widgets in current page, represented with text/id & [Component]: ``Depart, Arrive, Departure time, ...'' \\
 & [Position] & Relative position of widgets, obtained through their coordinates & [Position]: {Upper: ``Flight Search, ...''}, {Lower: ``Departure time, ... ''} \\
\midrule
\multirow{2}{2.2cm}{Input component information} & [TextInput] & The text input denoted with the textual related fields & [TextInput]:``Departure time'' \\
& [NearbyInput] & Nearby widgets denoted with their textual related fields & [NearbyInput] : ``Flight Search, ... '' \\

\bottomrule
\end{tabular}
\end{center}
\vspace{-0.05in}
\end{table*}

%% file: tab/example-prompt.tex
\begin{table*}[htb]
% \renewcommand\arraystretch{1.4} 
% \vspace{0.1in}
\caption{The example of the GUI prompt construction. It provides the app, GUI page and the input component information, then queries the LLM for the hint-text and its corresponding input content.}
\label{tab:example-prompt}
\vspace{-0.1in}
\centering
\footnotesize
\begin{center}
\begin{tabular}{m{0.6cm}<{\centering} | m{3.5cm}<{} | m{4.2cm}<{} | m{7.1cm}<{}}
\toprule
\textbf{Id} & \textbf{Prompt Type} & \textbf{Instantiation}  & \textbf{Examples} \\
\midrule
\multicolumn{4}{c}{\textbf{In-context learning prompt}} \\
\midrule
1 & <Hint-text examples> & We will provide you with 6 examples: 
 & We will provide you with 6 examples: \\
 &   & 1. [TextInput], [NearbyInput], [Hint-text] & \textbf{1st} text input is ``From'', its nearby components are ``...'', its hint-text is ... \\
 & & 2. [TextInput], [NearbyInput], [Hint-text] & \textbf{2nd} text input is ``To'', its nearby components are ..., its hint-text is ... \\
 & & ... & ...\\
 & & 6. [TextInput], [NearbyInput], [Hint-text] & \textbf{6th} text input is ``Flight'', its nearby ..., its hint-text is ``Enter the city''. \\
\midrule
\multicolumn{4}{c}{\textbf{GUI prompt}}\\ 
\midrule
2 & <App info> & [AppName], [Activities] & The app name is ``Flight'', it has following activities: ``Main, ...''  \\
3 & <Page GUI info> & [ActivityName], [Component], [Position] & The current GUI page is ``SearchFlight'', it has following components:``Search, ...'', the upper part of the page is ``...'', the lower part ....\\
4 & <Input component info> & [TextInput], [NearbyInput] & The text input of this page is ``Depart'', its nearby components are ... .\\
\midrule
\multicolumn{4}{c}{\textbf{Feedback prompt}} \\ 
\midrule
5 & <Feedback> & [Feedback], [ErrorMessage] & The input content ``train'' doesn't pass the page, the error message of the input component is: ``Please enter the correct city name''.\\
\midrule
\multicolumn{4}{c}{\textbf{Query \& feedback query}} \\ 
\midrule
6 & <Query> & \multicolumn{2}{l}{Please generate a hint-text for the input component based on the above information, and generate corresponding } \\ 
& & \multicolumn{2}{l}{input content based on the generated hint-text.} \\
7 & <Feedback Query> & \multicolumn{2}{l}{Please regenerate the hint text and its corresponding input content based on the feedback information above.} \\
\midrule
\multicolumn{4}{c}{\textbf{Example output}} \\ 
\midrule
8 & <Example output> & [Hint-text], [InputContent] & Please output according to the following example: the hint-text is ``xxx'', the input content is ``xxx''.   \\

\bottomrule
\end{tabular}
\end{center}
\vspace{-0.05in}
\end{table*}

%% file: figure/prompt-example.tex
\begin{figure*}[htb]
\centering
\vspace{0.1in}
\includegraphics[width=17.3cm]{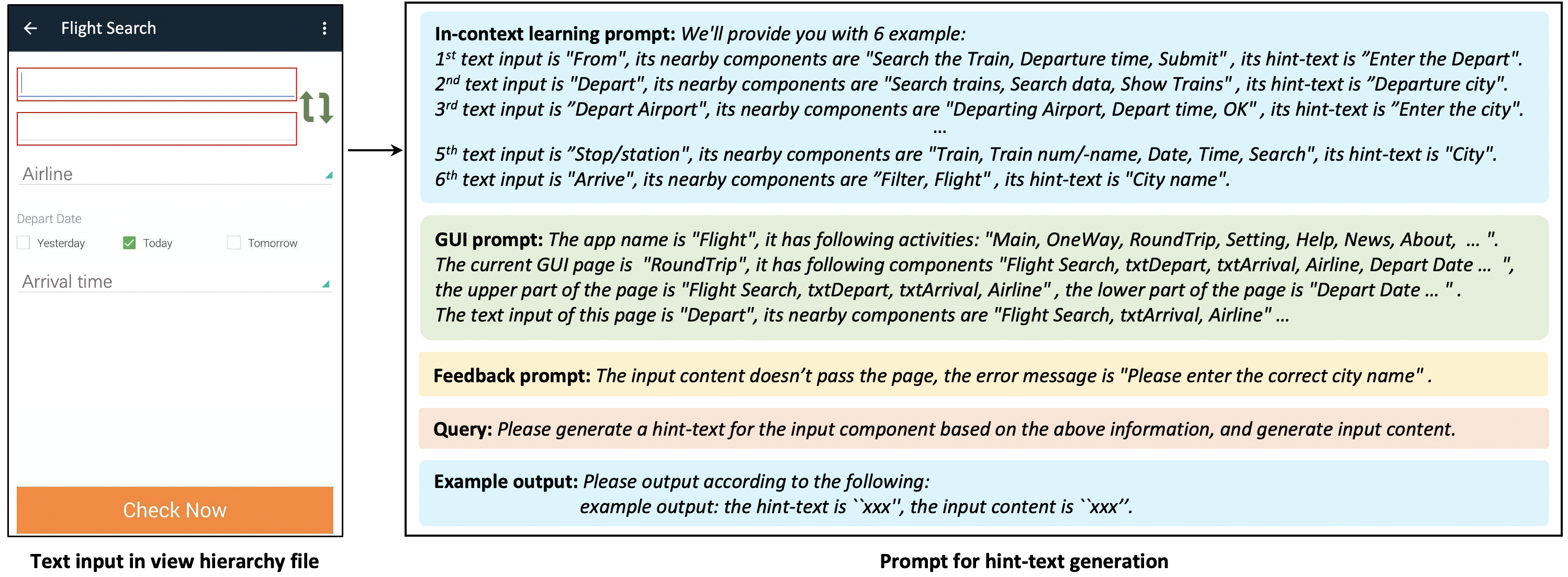}
\caption{Example of the prompt generation. The prompts include: in-context learning prompt, GUI prompt, feedback prompt, query and example output.}
\vspace{-0.1in}
\label{fig:prompt-example}
% \vspace{-0.1in}
\end{figure*}

%% file: figure/implement.tex
\begin{figure*}[htb]
\centering
% \vspace{0.1in}
\includegraphics[width=17.3cm]{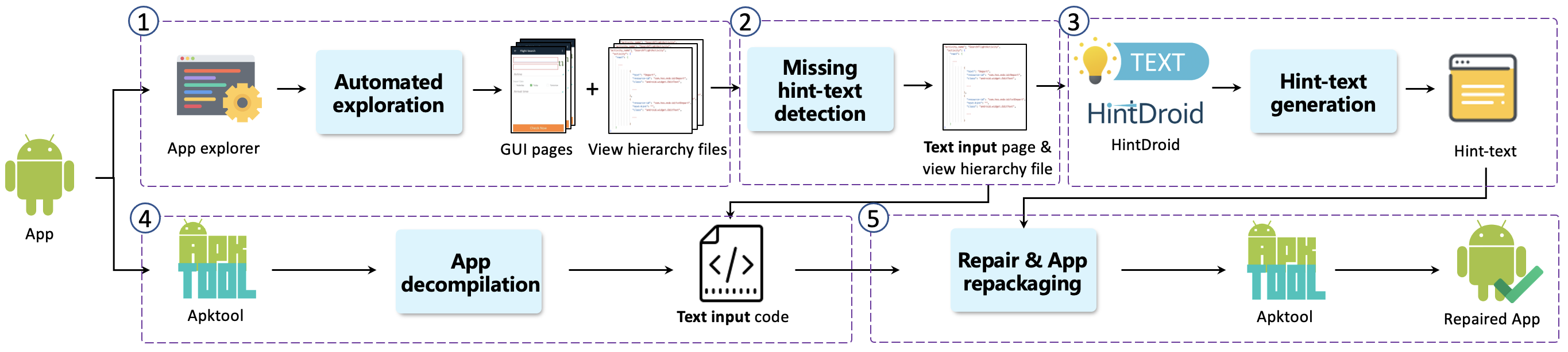}
\vspace{-0.1in}
\caption{Workflow of implementation. \ding{172} Extracting GUI pages. \ding{173} Detecting GUI pages with missing hint-text. \ding{174} Predicting hint-text based on GUI information. \ding{175} Decompiling APK to obtain code. \ding{176} Repackaging APK after code modification.}
\label{fig:implement}
\vspace{-0.1in}
\end{figure*}

%% file: sec/effectiveness.tex
\section{Effectiveness Evaluation}
\label{sec_Effectiveness}
We evaluate the effectiveness of {\tool} from the point of view of the hint-text generation accuracy.
For the accuracy of hint-text generation, we compare the exact match, BlEU, METEOR, ROUGE-L and CIDEr with 12 baseline methods to demonstrate its advantage (details are in Section \ref{subsec_experiment_baseline}).
For the model structure, we conduct ablation experiments to evaluate the impact of each (sub-) module on the performance.

\subsection{Experiment Setup}
\label{subsec_experiment_Setup}

\subsubsection{\textbf{Dataset and Experiment Procedures.} }
\label{subsec_experiment_dataset}
We collect the GUI pages that contain text input components and their corresponding hint-text from popular apps on Google Play as the experimental dataset. 
Specifically, we follow the data collection method in the motivational study in Section \ref{sec_motivation} and randomly select 50 apps from each of 33 categories that are different from the motivational study dataset, totaling 1,650 apps. We further ensure that the experimental data doesn't overlap with previous data through app name matching. We further use the Application Explorer~\cite{chen2018ui} to obtain GUI page file for each app, and filter out GUI pages with text input components and their corresponding hint-text through the ``hint'' attribute field. According to the above standards, we have obtained a total of 2,797 text input components from 753 apps with hint-text. We further recruited 2 developers with over 5 years of development experience to evaluate the quality of these hint-texts. Developers evaluate hint-text based on the accessibility specifications in the Google Developer Guidelines and annotate hint-text based on the principle of open coding protocol~\cite{seaman1999qualitative}.
The third developer reexamines the results until a consensus is reached.
In the end, 2,659 text input components with hint-text are used for our effectiveness evaluation.

\input{tab/RQ1}

\subsubsection{\textbf{Baselines}}
\label{subsec_experiment_baseline}
Since there are hardly any existing approaches for the hint-text generation of mobile apps, we employ 12 baselines from various aspects to provide a thorough comparison.

First, we directly utilize \textit{ChatGPT}~\cite{schulman2022chatgpt} as the baseline. We provide the GUI information of the text input component (as described in Table \ref{tab:example-prompt}), and ask it to generate hint-text.

Secondly, we use the hint-text example dataset constructed in Section \ref{subsubsec_approach_dataset_construction} to retrain the text generation model. The example dataset contains 15,577 pairs of GUI information for input components and their corresponding hint-text. We use GUI information as input to the model and hint-text as output to train the text generation model.
For text-based hint-text generation model, we select Recurrent Neural Network(RNN)~\cite{yin2017comparative}, LSTM~\cite{yu2019review}, Seq2Seq~\cite{li2018seq2seq}, Transformer~\cite{han2021transformer} as the hint-text generation models. 
For the image-based hint-text generation model, we choose Labeldroid~\cite{chen2020unblind} and CCN+LSTM~\cite{yin2017comparative}. Labeldroid~\cite{chen2020unblind} is a deep learning-based model to automatically predict the content description from the image button. Since our hint-text example dataset records GUI screenshots corresponding to hint-text, we also use this dataset as the training set to train these models.

Thirdly, considering that the input scenarios of some apps are similar to those in the example dataset, we design a retrieval-based matching method and a random-based matching method to select similar hint-text. We use the methods in Section \ref{subsubsec_approach_Retrieval} for information retrieval. Meanwhile, we also design a rule-based hint-text generation method, which has 36 general rules summarized by example data.

Fourthly, we select existing input text generation tools (QTypist and RNNInput) and use example data in Section \ref{subsubsec_approach_Retrieval} for fine-tuning. QTypist~\cite{liu2023fill} is a text input generation approach based on GPT-3, RNNInput~\cite{liu2017automatic} utilizes the RNN model and Word2Vec to predict the text input value for a given text input component.

\subsubsection{\textbf{Metric}}
\label{subsec_experiment_Metric}
To evaluate the performance of {\tool}, we select 5 widely-used evaluation metrics including exact match~\cite{chen2020unblind}, BLEU~\cite{papineni2002bleu}, METEOR~\cite{banerjee2005meteor}, ROUGE~\cite{lin2003automatic}, CIDEr~\cite{vedantam2015cider} inspired by related works about machine translation and image captioning. The exact match rate is the percentage of testing pairs whose predicted content description exactly matches the ground truth. Exact match is a binary metric, i.e., 0 if any difference, otherwise 1. It can't tell the extent to which a generated content description differs from the ground truth. Therefore, we also adopt other metrics. 

BLEU~\cite{papineni2002bleu} is an automatic evaluation metric widely used in machine translation. It calculates the similarity between machine-generated translations and human-created reference translations. BLEU is defined as the product of n-gram precision and brevity penalty. As most content descriptions for image-based buttons are short, we measure the BLEU value by setting n as 1, 2, 3, 4, represented as BLEU@1, BLEU@2, BLEU@3 and BLEU@4.

METEOR~\cite{banerjee2005meteor} (Metric for Evaluation of Translation with Explicit ORdering) is another metric used for machine translation evaluation. It is proposed to fix some disadvantages of BLEU which ignores the existence of synonyms and recall ratio. 
ROUGE (Recall-Oriented Understudy for Gisting Evaluation)~\cite{lin2003automatic} is a set of metrics based on recall rate, and we use ROUGE-L, which calculates the similarity between predicted sentence and reference based on the longest common subsequence.
CIDEr (Consensus-Based Image Description Evaluation)~\cite{vedantam2015cider} uses the term frequency-inverse document frequency to calculate the weights in reference sentences for different n-grams. 

All of these metrics give a real value with range [0,1] and are usually expressed as a percentage. The higher the metric score, the more similar the machine-generated content description is to the ground truth. If the predicted results exactly match the ground truth, the score of these metrics is 1 (100\%). We compute these metrics using coco-caption code~\cite{chen2015microsoft}.

\subsection{Results and Analysis}
\label{subsec_results}

\subsubsection{\textbf{Accuracy of Hint-text Generation}}
\label{sec_results_RQ1}
Table \ref{tab:RQ1} shows the overall hint-text generation performance of {\tool} and the baselines.
{\tool} achieves the average exact match, BLEU@1, BLEU@2, BLEU@3, BLEU@4, METEOR, ROUGE-L and CIDEr of 0.71, 0.83, 0.77, 0.73, 0.66, 0.67, 0.63, 0.62 across 2,659 hint-texts. This indicates the effectiveness of our approach in generating hint-text for text input components. 
We further examine the bad cases for the hint-text generated by {\tool} and summarize the following two reasons. 
(1) The GUI pages don't have contextual GUI components or the components without semantic information. In this case, observing the GUI context information of its previous page can enhance the understanding of the input component. (2) Hint-text expresses similar meanings, but it doesn't fully match the benchmark. We manually analyze these hint-texts and find that 93\% of them expressed similar meanings.

\input{figure/badcase}

To show the generalization of {\tool}, we also calculate the performance of {\tool} in different app categories as seen in Figure \ref{fig:RQ1}. We find that our proposed approach can generate diversified hint-text for different categories of apps which can help users understand the input requirements. Furthermore, it is good at capturing the contextual semantic information of the input components and generating the hint-text. {\tool} is also not sensitive to the app category, i.e., steady performance across different app categories.

\textbf{Performance comparison with baselines.} 
Table \ref{tab:RQ1} also shows the performance comparison with the baselines. 
We can see that our proposed {\tool} is much better than the baselines, i.e., 82\%, 0.57\%, 0.64\%, 0.70\%, 0.65\%, 0.43\%, 0.75\%, 0.77\% higher in exact match, BLEU@1, BLEU@2, BLEU@3, BLEU@4, METEOR, ROUGE-L and CIDEr compared with the best baseline Transformer.
We analyze the reasons for the failure of these baselines, mainly because they focus on generating correct input content for mobile GUI testing, rather than generating hint-text from the perspective of helping users and human cognition.
This further indicates the advantages of our approach.
We conduct the Mann-Whitney U test~\cite{mann1947test} between these models among all testing metrics. Since we have inferential statistical tests, we apply the Benjamini \& Hochberg (BH)~\cite{benjamini1995controlling} method to correct p-values. Results show the improvement of our model is significant in all comparisons (p-value<0.01).
Without our elaborate design, the raw ChatGPT demonstrates poor performance, which further indicates the necessity of our approach. 

\input{figure/RQ1}

\subsection{Ablation Study}
\label{sec_results_RQ2}

\subsubsection{\textbf{Contribution of Modules}}
\label{subsub_results_RQ2_Main}
Figure \ref{fig:RQ2} (a) shows the performance of {\tool} and its 2 variants respectively removing the second and third modules. 
In detail, for \textit{{\tool} w/o Module 2 (in-context learning prompt)}, we don't provide the example date for LLM. For \textit{{\tool} w/o Module 3 (Feedback prompt)}, we don't use feedback and just use the results generated once.
we provide the information related to the input components (as Table \ref{tab:example-prompt}) to the LLM.

We can see that {\tool}'s hint-text generation performance is much higher than all other variants, indicating the necessity of the designed modules and the advantage of our approach.
Compared with {\tool}, \textit{{\tool} w/o Module 3} results in the largest performance decline, i.e., 106\% drop (0.30 vs. 0.62) in CIDEr rate.
This further indicates that the feedback module can help LLM to deeply understand the input requirements and optimize the generated hint-text based on the error message.

\textit{{\tool} w/o in-context learning prompt} also undergoes a big performance decrease, i.e., 93\% (0.32 vs. 0.62) in CIDEr rate.
This might be because without the examples, the LLM would not understand the input intention and criteria for what kinds of hint-text are needed. 

\input{figure/RQ2}

\textbf{Contribution of Sub-modules.}
Figure \ref{fig:RQ2} further demonstrates the performance of {\tool} and its 4 variants. 
We remove the part of prompt when querying LLM, i.e., App information, page GUI information, input component information and error message. 
The experimental results demonstrate that removing any of the sub-modules would result in a noticeable performance decline, indicating the necessity and effectiveness of the designed sub-modules.

\subsubsection{\textbf{Influence of Different Number of Examples}}
\label{subsub_results_RQ2_Main_Example_data}
Figure \ref{fig:RQ2} demonstrates the performance under the different number of examples provided in the prompt. 
We can see that the hint-text generation performance increases with more examples, reaching the highest exact match, BLEU, METEOR, ROUGE-L and CIDEr with 6 examples.  
And after that, the performance would gradually decrease even increasing the examples. 
This indicates that too few or too many examples would both damage the performance, because of the tiny information or the noise in the provided examples.

%% file: tab/RQ1.tex
\begin{table*}[htb]
\vspace{0.1in}
\renewcommand\arraystretch{1.1} 
\caption{Result of the accuracy of hint-text generated by {\tool} and baselines.}
\vspace{-0.05in}
\label{tab:RQ1}
\centering
\footnotesize
\begin{tabular}{p{2.3cm}<{\centering} | p{2.1cm}<{\centering} | p{1.2cm}<{\centering} | p{1.2cm}<{\centering} | p{1.2cm}<{\centering} | p{1.2cm}<{\centering} | p{1.2cm}<{\centering} | p{1.2cm}<{\centering} | p{1.2cm}<{\centering} }
% \hline
\toprule
\textbf{Method} & \textbf{Exact match} & \textbf{BLEU@1} & \textbf{BLEU@2} & \textbf{BLEU@3} & \textbf{BLEU@4} & \textbf{METEOR} & \textbf{ROUGE-L} & \textbf{CIDEr}\\ 
\midrule
\multicolumn{9}{c}{\textbf{Learning-based method}} \\
\midrule
RNN & 0.29 & 0.37 & 0.35 & 0.31 & 0.29 & 0.26 & 0.24 & 0.21 \\
LSTM & 0.28 & 0.33 & 0.31 & 0.25 & 0.22 & 0.19 & 0.17 & 0.13 \\
Seq2Seq & 0.30 & 0.37 & 0.32 & 0.29 & 0.27 & 0.25 & 0.21 & 0.18 \\
Transformer & 0.39 & 0.53 & 0.47 & 0.43 & 0.40 & 0.37 & 0.36 & 0.35 \\
RNNInput & 0.28 & 0.35 & 0.33 & 0.32 & 0.27 & 0.25 & 0.23 & 0.19 \\
\midrule
LableDroid & 0.34 & 0.47 & 0.45 & 0.39 & 0.36 & 0.35 & 0.32 & 0.31 \\
CNN+LSTM & 0.26 & 0.29 & 0.24 & 0.19 & 0.17 & 0.15 & 0.09 & 0.08 \\
\midrule
\multicolumn{9}{c}{\textbf{Matching-based method}}\\
\midrule
Retrieval based & 0.21 & 0.27 & 0.24 & 0.22 & 0.18 & 0.20 & 0.18 & 0.15 \\
Random based & 0.11 & 0.16 & 0.13 & 0.10 & 0.07 & 0.08 & 0.09 & 0.07 \\
Rule based & 0.32 & 0.45 & 0.39 & 0.32 & 0.27 & 0.29 & 0.31 & 0.27 \\
\midrule
\multicolumn{9}{c}{\textbf{LLM-based method}}\\
\midrule
GhatGPT & 0.35 & 0.49 & 0.43 & 0.39 & 0.36 & 0.38 & 0.33 & 0.31 \\
QTypist & 0.31 & 0.47 & 0.41 & 0.35 & 0.33 & 0.37 & 0.34 & 0.33 \\
\midrule
{\tool} & \textbf{0.71} & \textbf{0.83} & \textbf{0.77} & \textbf{0.73} & \textbf{0.66} & \textbf{0.67} & \textbf{0.63} & \textbf{0.62} \\
\bottomrule
\end{tabular}
\vspace{-0.1in}
\end{table*}

%% file: figure/badcase.tex
\begin{figure*}[htb]
\centering
% \vspace{0.05in}
\includegraphics[width=15cm]{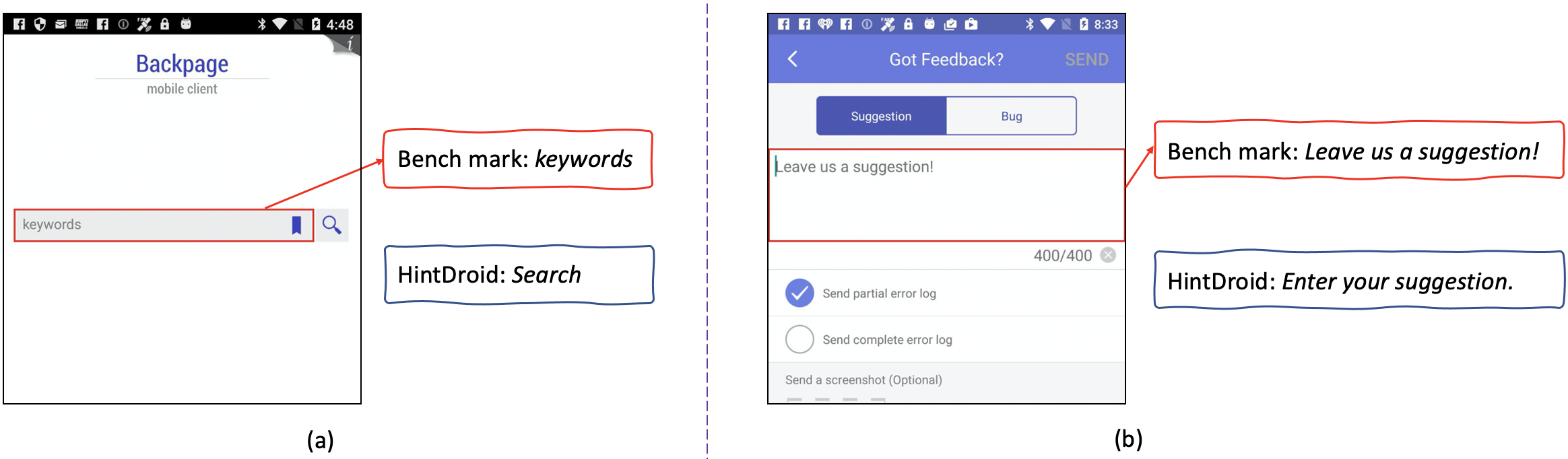}
\caption{Examples of bad case in hint-text generation. (a) The GUI pages don't have contextual GUI components or the components without semantic information. (b) Hint-text expresses similar meanings, but it doesn't fully match the benchmark.}
\vspace{-0.1in}
\label{fig:badcase}
\vspace{-0.05in}
\end{figure*}

%% file: figure/RQ1.tex
\begin{figure*}[htb]
\centering
\vspace{0.1in}
\includegraphics[width=17.3cm]{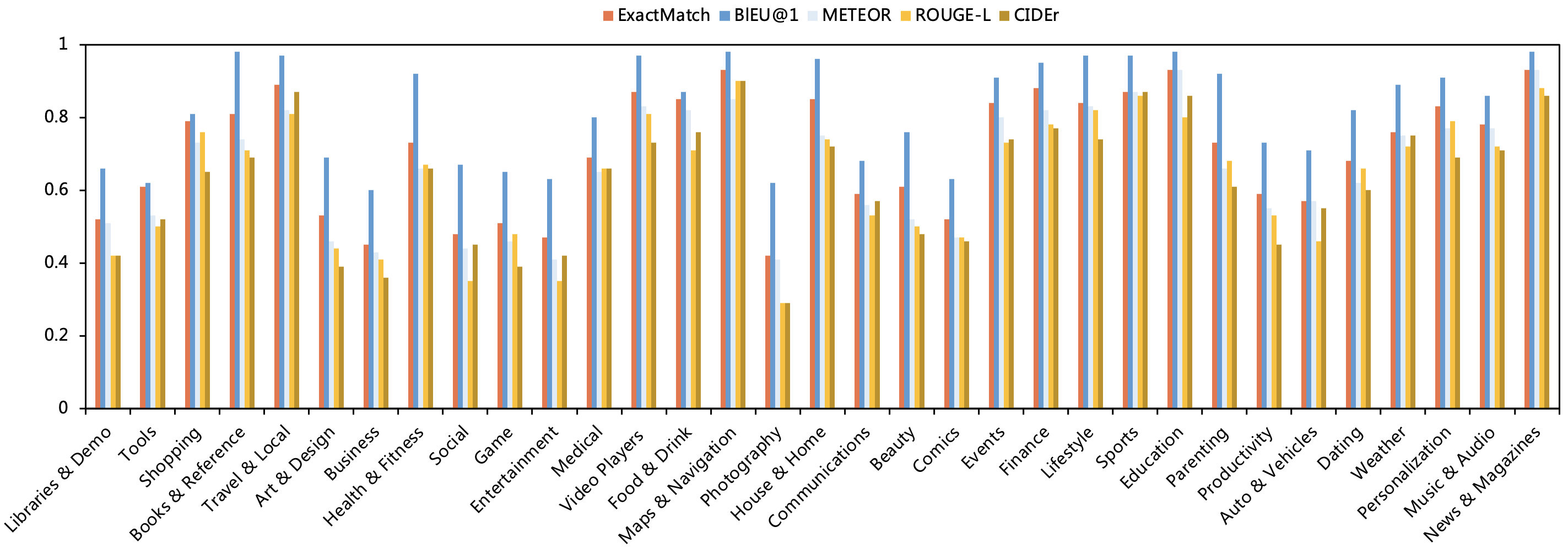}
\vspace{-0.1in}
\caption{Result of different app category. {\tool} can generate diversified hint-text for different categories of apps which can help users understand the input requirements.}
\label{fig:RQ1}
\vspace{-0.1in}
\end{figure*}

%% file: figure/RQ2.tex
\begin{figure*}[htb]
\centering
\vspace{0.1in}
\includegraphics[width=17.3cm]{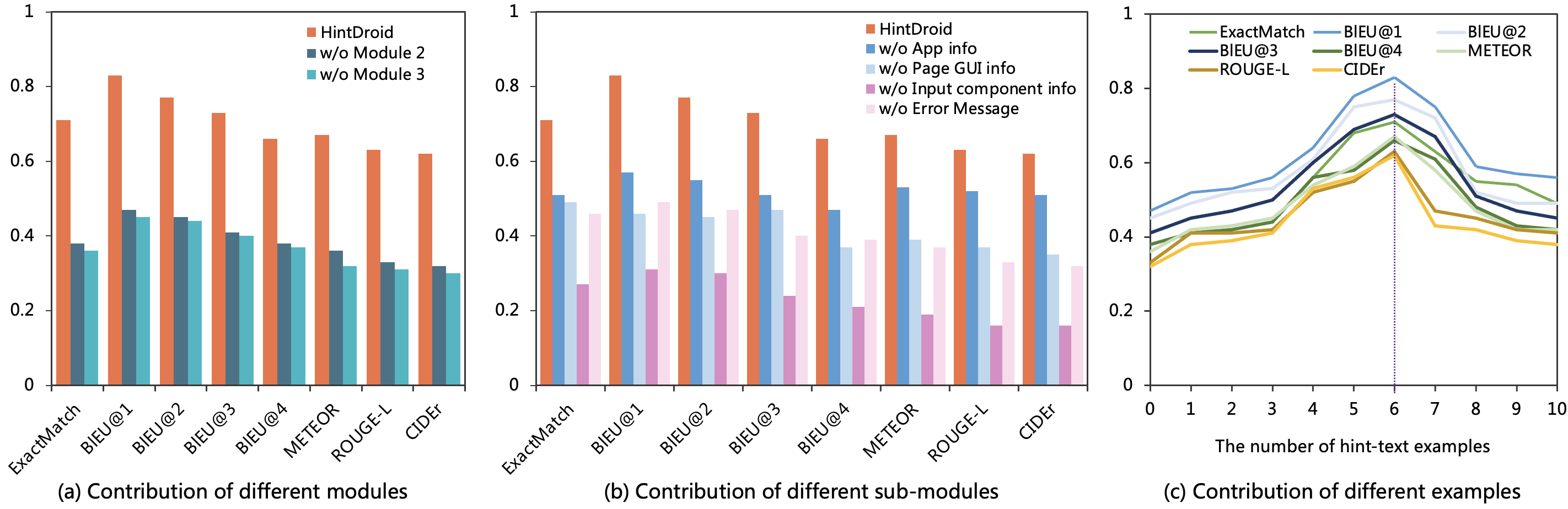}
\vspace{-0.1in}
\caption{Result of ablation study. The results demonstrate that removing any of the modules/sub-modules would result in a noticeable performance decline, indicating the necessity and effectiveness of the designed modules/sub-modules.}
\label{fig:RQ2}
\vspace{-0.1in}
\end{figure*}

%% file: sec/evaluation.tex
\section{Usefulness Evaluation}
\label{sec_Usefulness}
To evaluate our {\tool}, we also conduct a user study to demonstrate its usefulness in real-world practice. 
Our goal is to examine: (1) whether {\tool} can help visually impaired users successfully fill in the correct input.
(2) whether {\tool} can effectively help visually impaired users explore the functionality of the application. 
(3) whether {\tool} can save time by filling in the correct input.

\subsection{\textbf{Dataset of User Study}}
\label{sec_GUI_Testing_Tasks}
To ensure the representativeness of test data, We begin with the 3,398 apps from Google Play described in Section \ref{sec_motivation}, which have text input components without hint-texts (Please note that the data in this Section is not used for model training or in-context learning). To further confirm the universality and usefulness of our model, we first filter them according to the following rules. (1) At least 3 pages requiring text input components, (2) generated hint-texts can be integrated into the app, and (3) the number of activities in the app is more than 10. According to the above rules, we obtained 371 apps, and further selected the app with the highest download number of each app category as our experimental data. As shown in Figure \ref{fig:Dataset}, we describe the data selection process according to the PRISMA Flow Diagram\cite{page2021prisma}.
We end up with 33 apps (1 app per category) with 237 text input components, and use them for the final evaluation, with details in Table \ref{tab:RQ3}. 
\input{figure/Dataset}

\subsection{Participants Recruitment}
\label{sec_Participants}
We recruit 36 visually impaired users to participate in the experiment, of whom 20 are male and 16 are female. Their ages ranged from 20 to 55 years (median = 36 years). 22 participants have no residual vision, 8 have only light/dark perception and 6 have very little central vision. The participants have visual impairment for a period between 7 and 41 years. All participants use screen readers (Talkback~\cite{GoogleTalkBack}) as their primary assistive technology to use mobile apps. 
All of the participants have been using mobile devices for 5 years or more. Every participant receives a \$100 as a reward after the experiment. At the beginning of the experiment, we ask participants to use app functions as much as possible. We also conduct a follow-up survey among the participants regarding their experiment experience.

The study involves two groups of 36 participants: the experimental group from P1 to P18 who use the mobile apps with the hint-text generated by our {\tool}, and the control group from P19 to P36 who use the app without hint-text.
Each pair of participants $\langle$ Px, P(x+18) $\rangle$ has a comparable app using experience to ensure that the experimental group has similar expertise and capability to the control group in total~\cite{power2012guidelines,morris2018rich,choo2019examining,shinohara2022usability}.
Specifically, given that all participants are affiliated with the same rehabilitation and education institution, we seek the collaboration of the institution's director to assist in the matching process. This process ensures a one-to-one ratio between the experimental and control groups, with pairings based on comparable personal competencies. The director's extensive familiarity with the participants, stemming from over two years of close association, provided invaluable insight into their capabilities, ensuring an equitable and balanced distribution between the two study groups.

\subsection{Experimental Design}
\label{sec_Experimental_Design}
To avoid potential inconsistency, we pre-install the 36 apps in the Samsung Galaxy Note 10 with Android 9.0 OS. For each app in the experiment group, we first run Application Explorer~\cite{chen2018ui} to observe the GUI page file. Then run our {\tool} to complete the missing hint-text for each app. Finally, we repackage the APK file according to the automated script in Section \ref{subsec_approach_Implement}. Please note that to ensure the correctness of the experiment, we check that each app can still run correctly after repackaging.

We start the screen readers on the devices. and ask them to explore each app separately. 
The participants in the two groups need to use the 33 given mobile apps.
They are required to fully explore the app and cover as many functionalities as possible.
Each participant has up to 15 minutes to use a mobile app which is far more than the typical app session (71.56 seconds)~\cite{bohmer2011falling}. 
Each participant can choose whether to end the exploration early based on their perceived exploration situation.
Each of them conducts the experiment individually without any discussion with each other.
During their exploring, all their screen interactions are recorded, based on which we derive their exploring performance.

\subsection{Evaluation Metrics}
\label{sec_Coverage_Metrics}
Following previous studies~\cite{liu2023fill,li2017droidbot,chen2020improving}, we use the following metrics to evaluate the effectiveness of {\tool}.
\begin{itemize}
\item Input accuracy: (number of correct inputs filled in by user in an app) / (number of all input components in an app)
\item Activity coverage: (number of discovered activities) / (number of all activities)
\item State coverage: (number of discovered states) / (number of all possible states)
\item Filling time: average time spent from arriving at the page with text input to filling in the correct input.
\end{itemize}

\input{tab/RQ3}

\subsection{Results and Analysis}
\label{sec_Performance_Metrics}

We present the {\tool}'s input accuracy, average activity, state coverage and the average filling time across the two groups, as shown in Table~\ref{tab:RQ3}.

\subsubsection{\textbf{Higher Input Accuracy}}
\label{sec_coverage}
As shown in Table~\ref{tab:RQ3}, the average input accuracy of the experimental group is 0.83 which is about 152\% ((0.83-0.33)/0.33) higher than that of the control group.
The results of Mann-Whitney U Test~\cite{mann1947test} show there is a significant difference (p-value \textless 0.01) between these two groups in the input accuracy metrics.
This indicates that {\tool} can generate hint-text by analyzing the GUI information from text input components, helping visually impaired individuals better understand input requirements and successfully fill in the correct input. We also find that for some input components with limited information, the hint-text generated by the {\tool} is of great help to visually impaired individuals. 

\input{figure/RQ3}

We analyze these text input components and summarize them into three categories.
Firstly, some input components have content description/alt-text for abbreviations. For example, the abbreviation ``BFR'' for ``Body Fat Ratio'' as shown in Figure \ref{fig:RQ3} (a), commonly found in health apps, may not be understood by blind people.
Secondly, due to poor GUI design, there may only be simple icon components on the interface or colors to differentiate input. For example, Figure \ref{fig:RQ3} (b) shows an arrow icon for switching between departure city and arrival city and uses colors to differentiate them (visually impaired individuals can't distinguish colors), without providing a textual description of them.
Finally, some input components did not provide a hint and content description/alt-text, and it needs to be inferred based on the context. For example, as shown in Figure \ref{fig:RQ3} (c), in the date selection, users need to enter "diastolic pressure" and "systolic pressure" without any text explanation.

\subsubsection{\textbf{More Explored GUI Pages}}
\label{sec_results_RQ2_5}
With our {\tool}, the activity coverage of the experimental group is 0.69, which is about 77\% ((0.69-0.39)/0.39) higher than that of the control group. 
The state coverage of the experimental group is 0.73, which is about 66\% ((0.73-0.44)/0.44) higher than that of the control group. 
The results of Mann-Whitney U Test~\cite{mann1947test} shows there is a significant difference (p-value \textless 0.01, more detailed information of experiment can be seen in our website\textsuperscript{\ref{github}}.) between these two groups in both metrics.
It indicates that the hint-text generated by our {\tool} can help visually impaired users successfully fill in the correct input to explore more states and activities. We also find that {\tool} can help visually impaired users explore some activities that are hard to find without the help of the {\tool}. For example, some search-type text input components will not be able to access subsequent content if the correct search content is not entered.

\subsubsection{\textbf{Less Time Cost}}
\label{sec_results_RQ2_2}
It takes just 0.88 minutes for visually impaired users with our {\tool} to trigger the next page by filling in the correct input while 2.10 minutes in the control group.
The results of the Mann-Whitney U Test shows there are significant difference (p-value \textless 0.01) between these two groups for the filling time.
In fact, the average time of the control group is underestimated, because an average of 9 participants don't attempt to fill in the input or don't continue after filling in incorrect input, which means that they may need more time in this input function.
 
We watched the video recording of the app exploration in the control group to further discover the reasons for the higher time cost.
Without the {\tool}, we find that participants are unable to understand the requirements of the input component, attempting to input content based on one's own experience has a significant deviation from the actual input requirements. In contrast, participants who use the hint-text generated by {\tool} almost fill in the correct input in one go.
This observation further confirms the importance of the hint-text generated by our {\tool}.

\subsection{Users' Experience With {\tool}}
\label{sec_results_RQ2_3}
According to the visually impaired users' feedback, all of them confirm the usefulness of our {\tool} in assisting their app exploration.
They all appreciate that the hint-text generated by our {\tool} can help the understanding of the input requirement to successfully fill in the correct input, increasing the activity and state coverage.
For example, ``The hint-text generated by {\tool} is very helpful for us to fill in the input.''(P1),
``The hints were super helpful in guiding me through the input fields. Thanks for making it clear!''(P3), ``I really like the straightforward expression of these hints.''(P8), ``The hints gave me exactly what I needed to know.''(P14), ``They made it easy for me to fill in the blanks.''(P15).
Participants express they like the hint-text, such as ``Great, it's useful. I like it!''(P2), ``Yo, these hints were bang on!''(P5), ``Plain and simple!''(P9), ``These hints were crystal clear!''(P11), ``I liked how the hints were friendly.''(P12).
Participants express that our {\tool} can save their exploration time such as ``Nice job! The hint-text of {\tool} saves our time.''(P17), ``These hints made the whole input process so much faster.''(P13).

The participants also mention the drawback and potential improvement of our {\tool}.
They hope that we can also provide some examples of correct input or provide input formats (our method also can generate the input content based on hint-text). For example, ``If the hints could guide me better for specific formats like phone numbers or emails, that'd be awesome!''(P4), ``Providing links or references for more info in the hints would be really helpful.''(P7), ``Providing links or references for more info in the hints would be helpful.''(P10), ``If these prompts can show me how much I need to input, it might be more useful''(P18).
Participants also hope that {\tool} can be adjusted in real-time based on their input situation in the future.
For example, ``Can it be made into interactive hints? When I encounter problems, your tool can provide more details.''(P12), ``If we make a mistake, it’d be awesome if the hint could help us figure out what went wrong.''(P16).

%% file: figure/Dataset.tex
\begin{figure}[htb]
\centering
\vspace{0.1in}
\includegraphics[width=8.3cm]{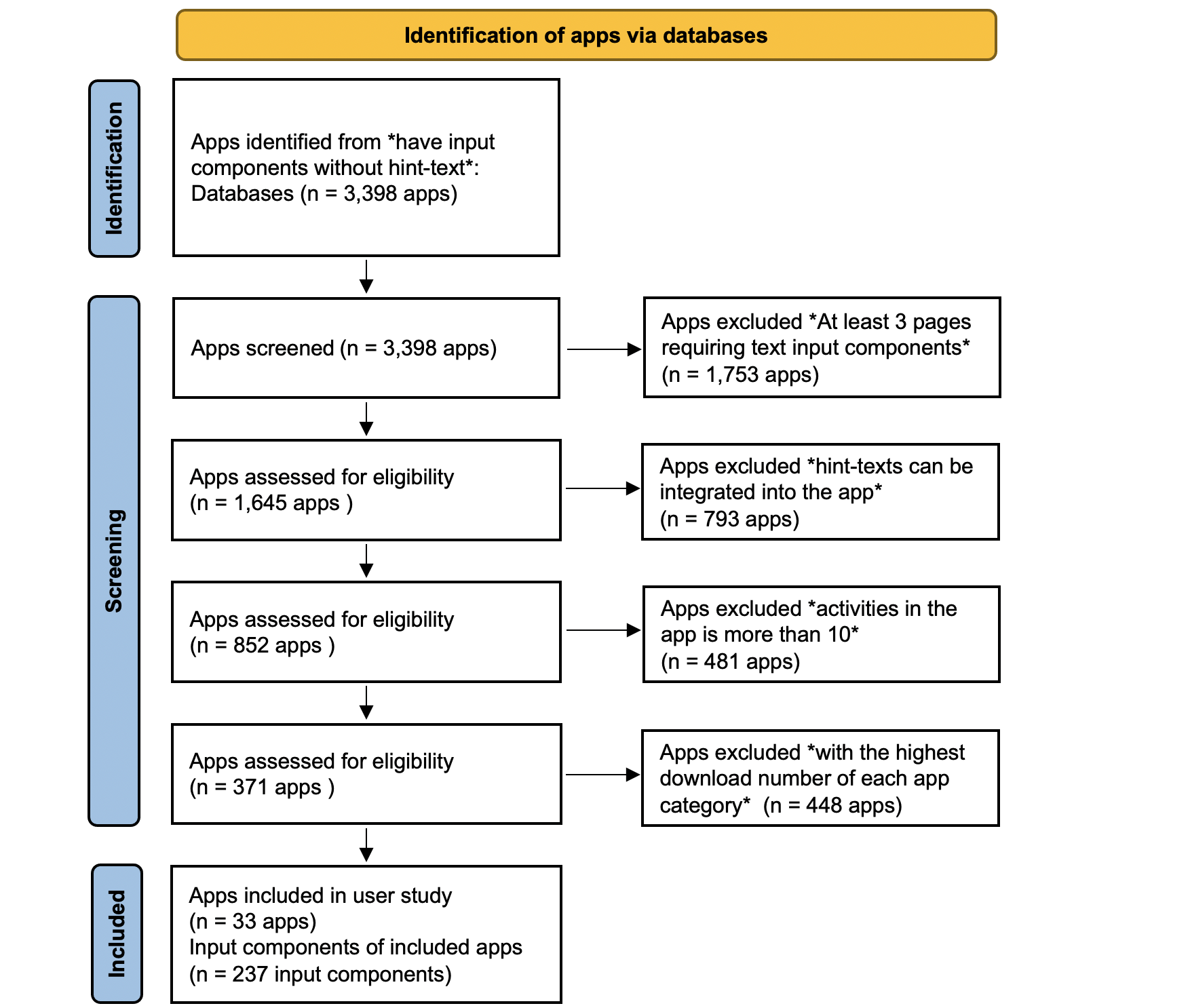}
\caption{Flowchart of data selection. We filter them according to 3 rules. (1) At least 3 pages requiring text input components, (2) generated hint-texts can be integrated into the app, and (3) the number of activities in the app is more than 10. }
\vspace{-0.1in}
\label{fig:Dataset}
\vspace{-0.05in}
\end{figure}

%% file: tab/RQ3.tex
\begin{table*}[htb]
\renewcommand\arraystretch{1.42} 
\vspace{0.2in}
\caption{The comparison of the experiment and control group. We present the {\tool}'s input accuracy, average activity, state coverage and the average filling time across the two groups.}
\label{tab:RQ3}
\centering
\footnotesize
\begin{tabular}{p{0.2cm}<{\centering}p{1.5cm}<{\centering}p{1.2cm}<{\centering}|p{1.1cm}<{\centering}p{1.5cm}<{\centering}|p{1.1cm}<{\centering}p{1.5cm}<{\centering}|p{1.1cm}<{\centering}p{1.5cm}<{\centering}|p{1.1cm}<{\centering}p{1.5cm}<{\centering}}
\toprule
\multicolumn{3}{c|}{\textbf{Basic information}} & \multicolumn{2}{c|}{\textbf{Input accuracy}} & \multicolumn{2}{c|}{\textbf{Activity coverage}} & \multicolumn{2}{c|}{\textbf{State coverage}} & \multicolumn{2}{c}{\textbf{Filling time (min)}} \\
% \cr\cline{1-11}
\midrule
 \textbf{id} & \textbf{App} & \textbf{Category} & \textbf{control} & \textbf{experiment} & \textbf{control} & \textbf{experiment} & \textbf{control} & \textbf{experiment} & \textbf{control} & \textbf{experiment}\\
\midrule
1 & HealthHB & Health & 0.50  & 0.81  & 0.54  & 0.72  & 0.51  & 0.64  & 2.38  & 1.18 \\ 
2 & WeatherL\&W & Weather & 0.40  & 0.88  & 0.24  & 0.66  & 0.62  & 0.78  & 1.93  & 1.02 \\ 
3 & MessagerWA & Communi & 0.31  & 0.95  & 0.29  & 0.73  & 0.24  & 0.68  & 2.30  & 1.10 \\ 
4 & MoneyTK & Finance & 0.53  & 0.72  & 0.50  & 0.69  & 0.36  & 0.74  & 2.74  & 0.70 \\ 
5 & FoodFacts & Food & 0.44  & 0.90  & 0.52  & 0.65  & 0.47  & 0.63  & 1.49  & 1.37 \\ 
6 & MPAS+ & Maps & 0.50  & 0.80  & 0.28  & 0.67  & 0.35  & 0.74  & 1.79  & 1.46 \\ 
7 & PSStore & Product & 0.34  & 0.80  & 0.28  & 0.63  & 0.52  & 0.76  & 1.50  & 1.43 \\ 
8 & NewAudio & Music & 0.42  & 0.84  & 0.20  & 0.64  & 0.51  & 0.82  & 2.89  & 0.64 \\ 
9 & WallETH & Personal & 0.47  & 0.90  & 0.21  & 0.74  & 0.26  & 0.81  & 1.72  & 0.93 \\ 
10 & PicGall & Photo & 0.45  & 0.71  & 0.27  & 0.59  & 0.37  & 0.81  & 2.92  & 0.53 \\ 
11 & SmartNew & News & 0.36  & 0.75  & 0.52  & 0.68  & 0.25  & 0.64  & 1.60  & 1.49 \\ 
12 & MyHM & House & 0.18  & 0.87  & 0.47  & 0.75  & 0.43  & 0.71  & 2.56  & 0.83 \\ 
13 & INSTEAD & Life & 0.15  & 0.77  & 0.53  & 0.69  & 0.30  & 0.75  & 1.80  & 0.54 \\ 
14 & GameSpe & Game & 0.55  & 0.73  & 0.25  & 0.75  & 0.50  & 0.76  & 1.71  & 0.51 \\ 
15 & BusinessEX & Business & 0.12  & 0.72  & 0.21  & 0.65  & 0.35  & 0.79  & 1.50  & 0.55 \\ 
16 & PocketMaps & Travel & 0.14  & 0.80  & 0.41  & 0.71  & 0.46  & 0.78  & 2.84  & 0.58 \\ 
17 & EventOR & Events & 0.12  & 0.95  & 0.52  & 0.73  & 0.57  & 0.75  & 1.43  & 0.23 \\ 
18 & FitTAP & Comics & 0.16  & 0.93  & 0.46  & 0.69  & 0.31  & 0.82  & 2.13  & 0.46 \\ 
19 & SkyTube & Video & 0.17  & 0.96  & 0.45  & 0.72  & 0.36  & 0.82  & 2.19  & 0.95 \\ 
20 & LibReader & Books & 0.26  & 0.71  & 0.33  & 0.69  & 0.25  & 0.69  & 2.50  & 0.32 \\ 
21 & NoxSecu & Tool & 0.53  & 0.96  & 0.47  & 0.68  & 0.63  & 0.64  & 2.03  & 1.29 \\ 
22 & EarnMon & Social & 0.13  & 0.91  & 0.39  & 0.73  & 0.63  & 0.66  & 2.29  & 0.32 \\ 
23 & WalkTra & Sports & 0.35  & 0.70  & 0.42  & 0.72  & 0.56  & 0.65  & 2.71  & 0.68 \\ 
24 & ParentLA & Parenting & 0.36  & 0.72  & 0.24  & 0.69  & 0.59  & 0.80  & 1.70  & 1.55 \\ 
25 & ISAY & Medical & 0.53  & 0.96  & 0.48  & 0.60  & 0.29  & 0.72  & 2.18  & 0.43 \\ 
26 & Ipsos & Commun & 0.09  & 0.72  & 0.54  & 0.79  & 0.55  & 0.79  & 2.71  & 1.25 \\ 
27 & FIRR & Libraries & 0.53  & 0.79  & 0.52  & 0.65  & 0.34  & 0.67  & 1.33  & 1.23 \\ 
28 & DRBUs & Shopping & 0.28  & 0.72  & 0.53  & 0.74  & 0.63  & 0.71  & 2.19  & 1.44 \\ 
29 & Learning & Education & 0.49  & 0.89  & 0.33  & 0.69  & 0.40  & 0.63  & 2.17  & 0.95 \\ 
30 & MMDR & Dating & 0.06  & 0.74  & 0.21  & 0.75  & 0.31  & 0.77  & 1.37  & 0.73 \\ 
31 & Pretty & Beauty & 0.19  & 0.94  & 0.58  & 0.70  & 0.60  & 0.73  & 1.99  & 0.36 \\ 
32 & Fair & Auto & 0.33  & 0.82  & 0.46  & 0.74  & 0.51  & 0.65  & 1.77  & 1.25 \\ 
33 & ArtPIX & Art & 0.49  & 0.85  & 0.24  & 0.57  & 0.41  & 0.67  & 2.88  & 0.86 \\ 
\midrule
\multicolumn{3}{c|}{\textbf{Average}} &  0.33  & \textbf{0.83}  & 0.39  & \textbf{0.69}  & 0.44  & \textbf{0.73}  & 2.10  & \textbf{0.88} \\ 
\bottomrule
\end{tabular}
% \vspace{-0.05in}
\end{table*}

%% file: figure/RQ3.tex
\begin{figure*}[htb]
\centering
\vspace{0.15in}
\includegraphics[width=17.3cm]{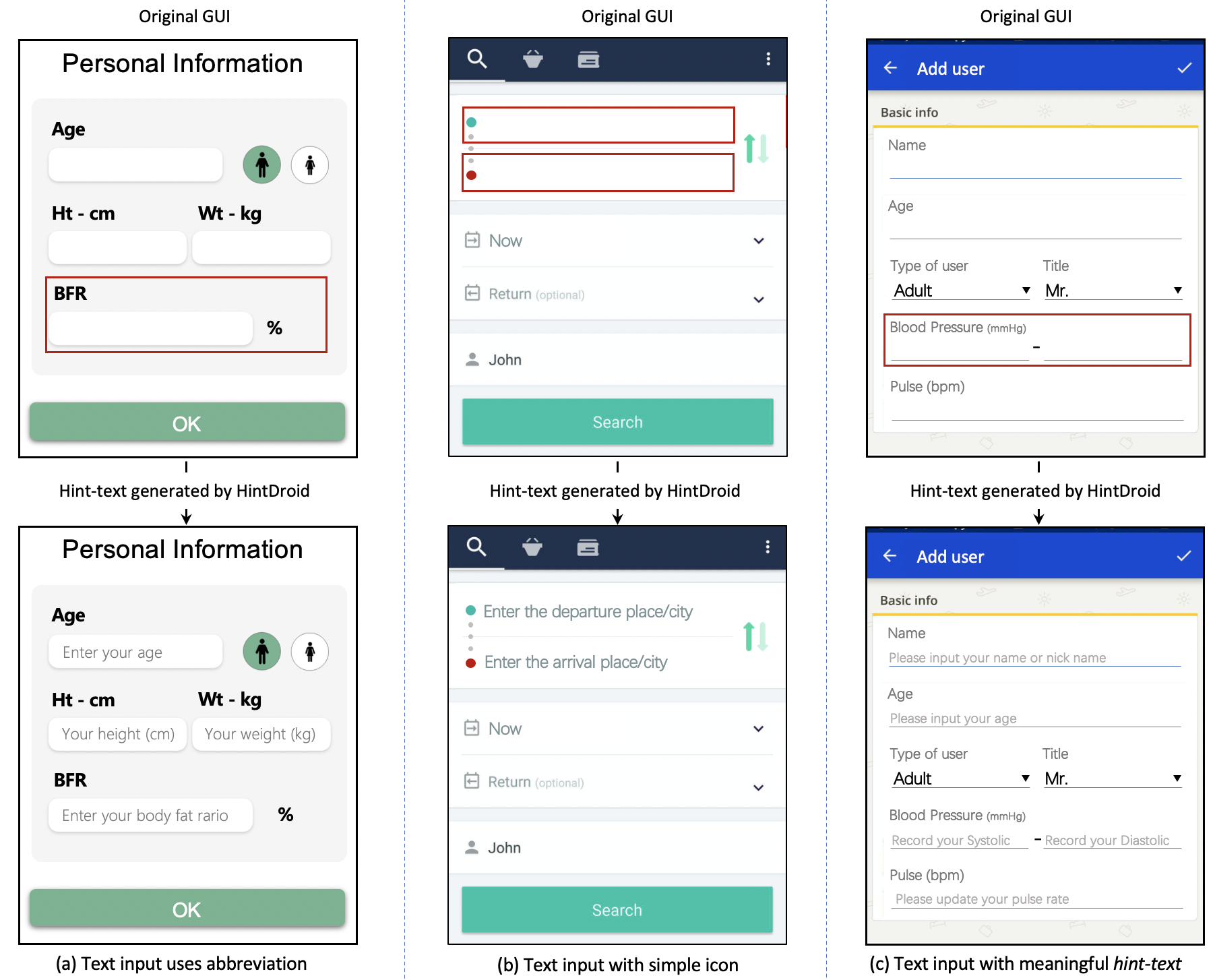}
\vspace{-0.1in}
\caption{Example of different good cases generated by {\tool}. (a) {\tool} generates the full name of the abbreviation and further explains it. (b) {\tool} provides explanations for the function of distinguishing by color. (c) {\tool} infers hint-text based on context.}
\label{fig:RQ3}
% \vspace{-0.1in}
\end{figure*}

%% file: sec/discussion.tex
\section{Discussion}
\label{sec_discussion}

In summary, we find that the hint-text generated by our {\tool} can effectively help visually impaired users successfully fill in the correct input.

\subsection{The Generalization of Our Approach and Findings}
{\tool} is designed to generate the hint-text of text input component, which can help visually impaired users successfully fill in the correct input. 
In addition to Android, there are also many other platforms such as iOS, Web and Desktop. To conquer the market, developers tend to develop either one cross-platform app or separate native apps for each platform considering the performance benefit of native apps.
Although our {\tool} is designed specifically for Android, since other platforms have similar types of information, it can also be extended to other platforms.

We conduct a small-scale experiment for another two popular platforms, and experiment on 20 iOS apps with 34 text input and 20 Web apps with 57 text input, with details on our website.
Results show that {\tool} achieves the average exact match, BLEU@1, of 0.73, 0.88 for iOS apps and 0.71, 0.85 for Web apps.
This further demonstrates the generality and usefulness of {\tool}, and we will conduct more thorough experiments in the future.

\subsection{Potential Applications to End-users}
In addition to generating the hint-text in helping the visually impaired users successfully fill in the correct input, our {\tool} can also be applied to help end-users in their daily app usage.
Given the increasing complexity of mobile apps, filling in the correct input of mobile apps is a challenging task, especially for aged users.
For example, there may be too many text input components in one GUI page for aged users to try the correct input.
They may stuck on one page with repetitive attempts but not working.
Even normal users may stay on the input page due to unclear input requirements(without hint-text), especially for new apps.

Based on {\tool}, according to the GUI information of the text input component and knowledge from popular app datasets, the {\tool} can automatically generate hint-text and complete the ``hint'' fields of the text input component in the app. Therefore, {\tool} can also be integrated into UI automation tools~\cite{hao2014puma,wen2023droidbot,krosnick2022parammacros} to provide developers with more diverse hint-text. In addition, {\tool} generates corresponding input content based on the generated hint-text, which can be used in software testing to help testers generate diverse test cases.

\subsection{Potential Directions to Improve the Hint-text}
Our experiment shows that hint-text can help visually impaired individuals understand input needs. However, developers lack a unified style and approach when designing hint-texts, which may lead to ambiguity for visually impaired individuals. {\tool} utilizes generative models to generate hint-text, which could potentially optimize this process in the future. For example, a hint-text generation model can be customized based on the historical usage records of visually impaired individuals, or a generative model can be customized based on the usage scenarios of different types of applications, providing personalized hint-text.

\subsection{Limitations}
\label{subsec_Limitations}
Although the average metric of the hint-text generated by {\tool} exceeds 70\%, there are still some inaccuracies in the generation of hint-texts. As analyzed in Section~\ref{sec_results_RQ1}, different developers have different design styles when designing text-input components, some of them have little or no contextual information. All these could influence the correct generation of hint-text. We will keep improving {\tool} for generating hint-text more accurately through the information from the previous GUI page.

For the correctness and rationality of the hint-text generated by {\tool}, we only consider whether the input content generated based on the hint-text can trigger page transitions. As described in Section \ref{subsec_approach_FeedBack}, not triggering page transitions is just the worst-case scenario. In addition, factors such as whether the information conveyed by hint-text is reasonable, complete, and ambiguous need to be considered. We will also incorporate these evaluation indicators in our future work.

In addition, {\tool} is currently an offline one-time automated hint-text missing issue repair approach that uses repackaging to complete hint-text injection. For some closed-source apps that use code encryption, code obfuscation, and other techniques that prevent decompilation and repackaging, we will also send the hint-text generated by {\tool} to the developers via email. Considering the low cost of the approach, i.e., the average time to generate hint-text on each GUI page is 1.86 seconds. Meanwhile, adopting repackaging technology may potentially bring some security risks, so we prefer to implement our {\tool} through real-time interaction. As suggested by the participants in \ref{sec_results_RQ2_3}, we will design a real-time interaction approach in the future and integrate it into the screen reader. It is possible to dynamically adjust the description of hint-text based on user input.

%% file: sec/conclusion.tex
\section{Conclusion}
\label{sec_conclusion}
The development of applications brings a lot of convenience to the daily lives of visually impaired people. They can use screen readers embedded in mobile operating systems to read the content of each screen within the app and understand the content that needs to be operated. However, the issue of missing hint-text in the text input component poses a challenge for screen readers to obtain input information. Based on our analysis of 4,501 Android apps with text inputs, over 76\% of them are with missing hint-text. To overcome these challenges, we develop an LLM-based hint-text generation model called {\tool}, which analyzes the GUI information of input components and uses in-context learning to generate the hint-text. To ensure the quality of hint-text generation, we further design a feedback-based inspection mechanism to further optimize hint-text.
The automated experiments demonstrate the high BLEU and a user study further confirms its usefulness.

In the future, we will work in two directions.
First, we will improve the performance of our approach by extracting more GUI context information. 
According to the user feedback, we will optimize the hint-text generated by our {\tool}, which can borrow the idea from the human-machine collaboration studies to better provide convenience for users.
Second, we will not limit our {\tool} to assisting visually impaired individuals in app usage, and plan to explore its potential applications in the field of software development, such as integrating it into IDE.